\documentclass{ws-procs9x6}

\newcommand\sqrtsnn{\sqrt{s_{_{NN}}}}
\newcommand\pT{p_{_{T}}}
\newcommand\AuAu{${\rm Au} {\rm Au}$}
\newcommand\PbPb{${\rm Pb} {\rm Pb}$}
\newcommand\ee{${\rm e}^+ {\rm e}^-$}
\newcommand\pp{${\rm p} {\rm p}$}
\newcommand\pA{${\rm p} {\rm A}$}
\newcommand\ppbar{${\rm p} \bar{\rm p}$}
\newcommand\dAu{${\rm d} {\rm Au}$}
\newcommand\avgNp{\langle N_{part} \rangle}

\begin{document}

\title{Exploring QCD with Heavy Ion Collisions}

\author{M.~D. BAKER}

\address{Brookhaven National Laboratory, \\
        Bldg. 555A, P.O. Box 5000 \\
        Upton, NY 11973-5000, USA \\
	E-mail: Mark.Baker@bnl.gov}


\maketitle

\abstracts{
After decades of painstaking research, the field of heavy ion physics
has reached an exciting new era. Evidence is mounting that we can
create a high temperature, high density, strongly interacting ``bulk
matter'' state in the laboratory --- perhaps even a quark-gluon
plasma.  This strongly interacting matter is likely to provide
qualitative new information about the fundamental strong interaction,
described by Quantum Chromodynamics (QCD). These lectures provide
a summary of experimental heavy ion research, with particular
emphasis on recent results from RHIC (Relativistic Heavy Ion Collider)
at Brookhaven National Laboratory. In addition, we will discuss what
has been learned so far and the outstanding puzzles.}

\section{Introduction}

 
While the universe as we know it is well described by the standard
model of particle physics, some important questions remain unanswered.
Perturbative Quantum Chromodynamics (pQCD) --- a part of the standard
model --- is a very successful description of hard, or short-distance,
phenomena~\cite{QCDtome}, where the ``strong interaction'' becomes
weak due to asymptotic freedom. For example, the production of jets in
\ppbar\ collisions at 1.8~TeV is well described for jet transverse
energies from 10--400 GeV~\cite{CDFjets}.  There is, however, an
important set of soft physics phenomena that are not well
understood from first principles in QCD: color confinement, chiral
symmetry breaking, and the structure of the vacuum. These phenomena
are important: almost all of the visible mass of the universe is
generated by soft QCD and not by the direct Higgs mechanism.  The
current masses of the three valence quarks make up only about 1\% of
the mass of the nucleon~\cite{PDG}.

In order to study these phenomenon, we seek to separate color charges
by heating matter until a quark-gluon plasma is formed. A
conventional electromagnetic plasma occurs at temperatures of about
$10^4$--$10^5$~K, corresponding to the typical ionization energy scale
of 1--10~eV.  Theoretical studies of QCD on the lattice~\cite{Karsch}
indicate that the typical energy scales of thermally driven color
deconfinement are in the vicinity of 170~MeV, or $2 \times
10^{12}$ K. In addition to providing information about the strong
interaction, achieving such temperatures would also provide a window
back in time. The color confinement phase transition is believed to
have occurred within the first few microseconds after the big bang.

In order to achieve such high temperatures under laboratory
conditions, it is necessary to use a small, dynamic system. For
instance, experimental fusion reactors heat a conventional plasma up
to temperatures as high as $10^8$~K over distance scales of meters and
lasting for seconds.  By colliding gold ions at nearly the speed of
light, we expect to achieve temperatures of order $10^{12}$~K over
distance scales of order 10~fm and time scales of order
10--100~ys\footnote{Recall that one yoctosecond =
$10^{-24}$~s.}. Clearly one of the challenges in this endeavor will be
to determine whether such small and rapidly evolving systems can
elucidate the bulk behavior that we are interested in. Another
challenge will be to use some of the rarer products of the collisions
to probe the created ``bulk'' medium.

The focus of these lectures will be on the results coming out of the
Relativistic Heavy Ion Collider~(RHIC) experiments at Brookhaven
National Laboratory~(BNL). Earlier experimental results and some
theoretical work will be mentioned as needed, but a comprehensive
review of heavy ion physics will not be attempted. The RHIC spin
physics program using polarized protons will also not be covered.

\section{The Machine and Detectors}

The RHIC data described in these lectures were taken during the last
three years of running at RHIC, starting in the summer of 2000, as
summarized in Table~\ref{tab:RHICruns}. The runs were characterized
by their species and their $\sqrtsnn$, which is the cm collision
energy of one nucleon taken from each nucleus. For instance, a 
\AuAu\ collision with $100\times A$~GeV on $100\times A$~GeV would
have $\sqrtsnn=200$~GeV. Most of the runs were several
weeks in duration, with two exceptions. The 56~GeV run, not intended
as a physics run, was only 3 hours long and data is only available
from a preliminary subsystem of one experiment (PHOBOS). The 19.6~GeV
run was 24 hours long and usable data were taken by three
experiments. For the 130 and 200~GeV runs, all four detectors 
participated: two large detectors/collaborations with 
300--400 collaborators each and two small detectors/collaborations
with 50--70 collaborators each. These four detectors complement
each other and have provided a broad range of physics results. The
BRAHMS experiment (Broad RAnge Hadron Magnetic Spectrometer) focuses
on tracking and particle ID at high transverse momentum over a broad
range of rapidity from 0--3.  The PHENIX experiment (Pioneering High
Energy Nuclear Interaction eXperiment) provides a window primarily at
mid-rapidity, but specializes in high rate and sophisticated
triggering along with a capability to measure leptons and photons as
well as hadrons. The PHOBOS experiment (descendant of the earlier MARS
experiment) provides nearly 4$\pi$ coverage for charged particle
detection, good vertex resolution, and sensitivity to very low $\pT$
particles. The STAR experiment (Solenoidal Tracker At
Rhic) provides large solid angle tracking and complete coverage of
every event written to tape. More details concerning the capabilities
of the accelerator and experiments can be found in NIM journal issue
dedicated to the RHIC accelerator and detectors~\cite{NIMRHIC}.

\begin{table}[ph]
\tbl{RHIC running conditions to date.\vspace*{1pt}}
{
\begin{tabular}{|c|cl|}
\hline
Date & Species & $\sqrtsnn$ \\[0.2ex]
\hline
Year 2000 & \AuAu     & 56, 130 GeV \\[1ex]
Year 2001 & \AuAu     & 19.6, 200 GeV \\[1ex]
Jan. 2002 & \pp       & 200 GeV         \\[1ex]
Year 2003 & \dAu, \pp & 200 GeV\\[1ex]
\hline
\end{tabular}\label{tab:RHICruns} }
\vspace*{-13pt}
\end{table}

Some data will also be shown from lower energy heavy ion collisions,
particularly from the CERN-SPS (Conseil European pour la Recherch\'e
Nuclearie - Super Proton Synchrotron) will also be discussed. The top
CERN energy is $\sqrtsnn=17.2$~GeV.

\section{Strongly Interacting Bulk Matter}

In order to learn anything about QCD from heavy ion collisions, we
must first establish that we have created a state of strongly
interacting bulk matter under extreme conditions of temperature and
pressure.

\subsection{How Much Matter?}

\begin{figure}[htbp]
\centerline{
\epsfxsize=8cm\epsfbox{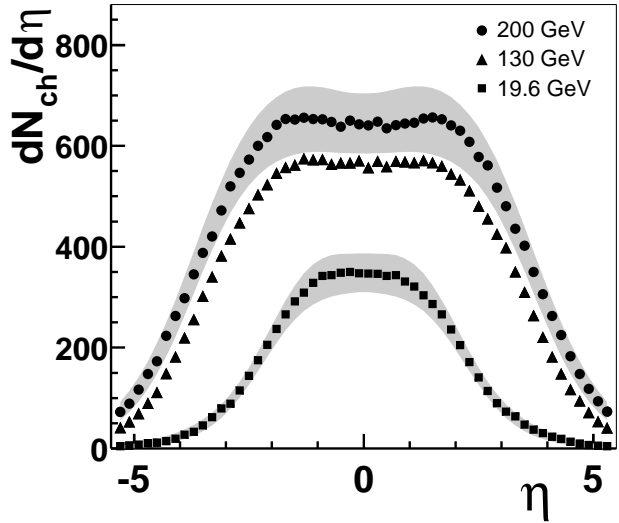}}
\caption{Pseudorapidity distributions, $dN_{ch}/d\eta$, for central
(6\%) \AuAu\ collisions at $\sqrt{s_{_{NN}}} =$ 19.6, 130, and
200~GeV.  Data taken from PHOBOS~\protect\cite{PHO:limfragprl}.}
\label{fig:Pho3energies}
\end{figure}
Figure~\ref{fig:Pho3energies} shows the charged particle distribution
for central (head-on) \AuAu\ collisions in the pseudorapidity
variable: $\eta \equiv - \ln \tan (\theta/2)$. These data imply a
total charged multiplicity of $1680 \pm 100$ for the 19.6~GeV data and
$5060 \pm 250$ for the 200~GeV data~\cite{PHO:limfragprl}. While this
number is considerably smaller than Avagadro's number, it is
substantial thermodynamically since small-system corrections to
conventional thermodynamics start to become unimportant for systems
with about 1000 particles or more~\cite{Hill:SmallSystems}.

The number of particles produced in a given \AuAu\ collision
varies widely due to the variable geometry of the collision. Some
collisions are nearly head-on with a small impact parameter,
while most collisions have a larger impact parameter, with only a
partial overlap of the nuclei. These cases can be sorted out
experimentally, using both the number of produced particles and the
amount of ``spectator'' neutrons seen at nearly zero degrees along the
beam axis. The impact parameter or ``centrality'' of the collision is
characterized by the number of nucleons from the original ions which
participate in the heavy ion collision, $\langle N_{part} \rangle$, or
the number of binary NN collisions, $\langle N_{coll} \rangle$. More
details can be found in Refs.~\cite{PHX:caterpillar,PHO:cent}.

\subsection{Elliptic Flow: Evidence for Collective Motion}

\begin{figure}[htbp]
\centerline{
\epsfxsize=72mm\epsfbox{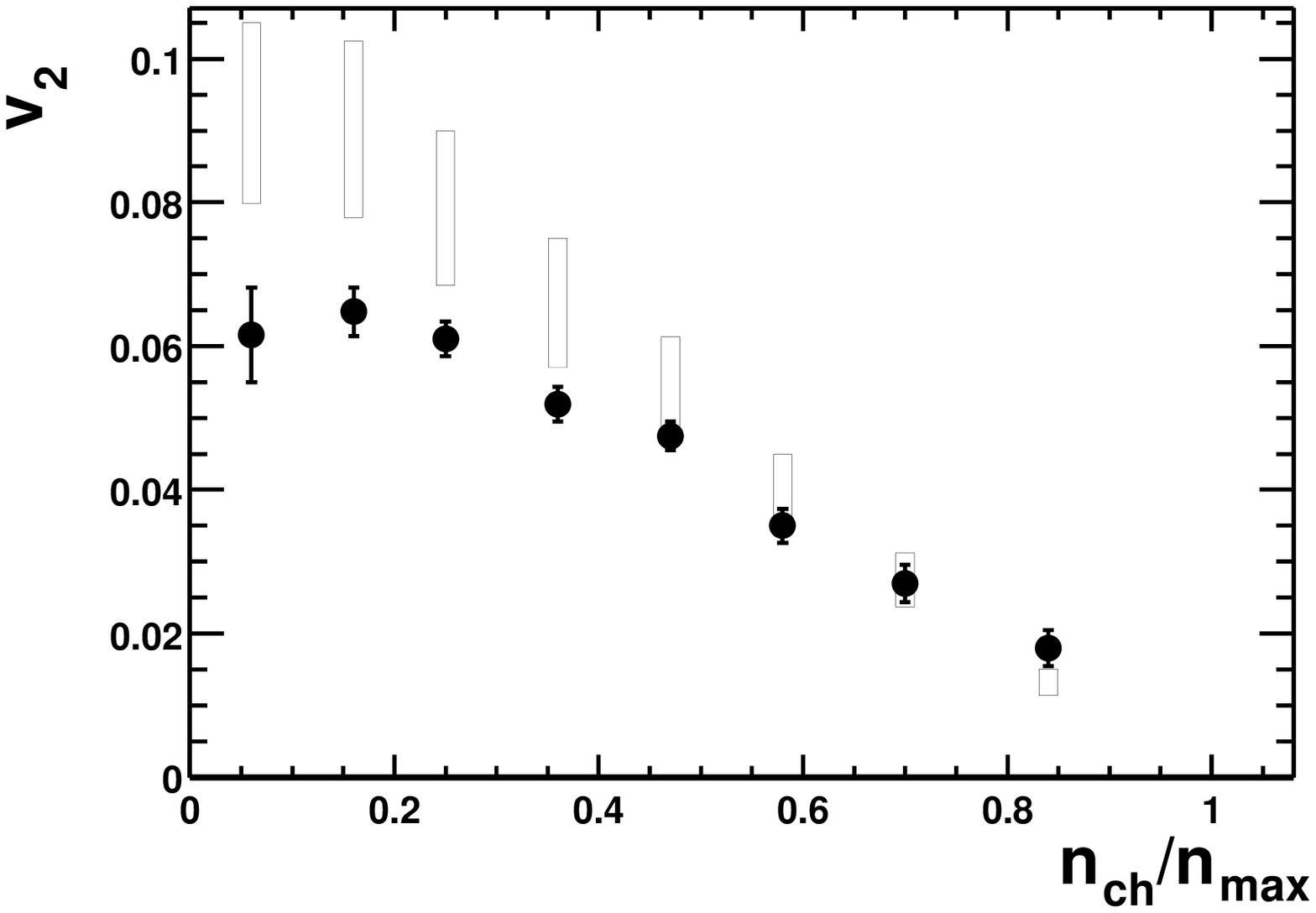}
\epsfxsize=48mm\epsfbox{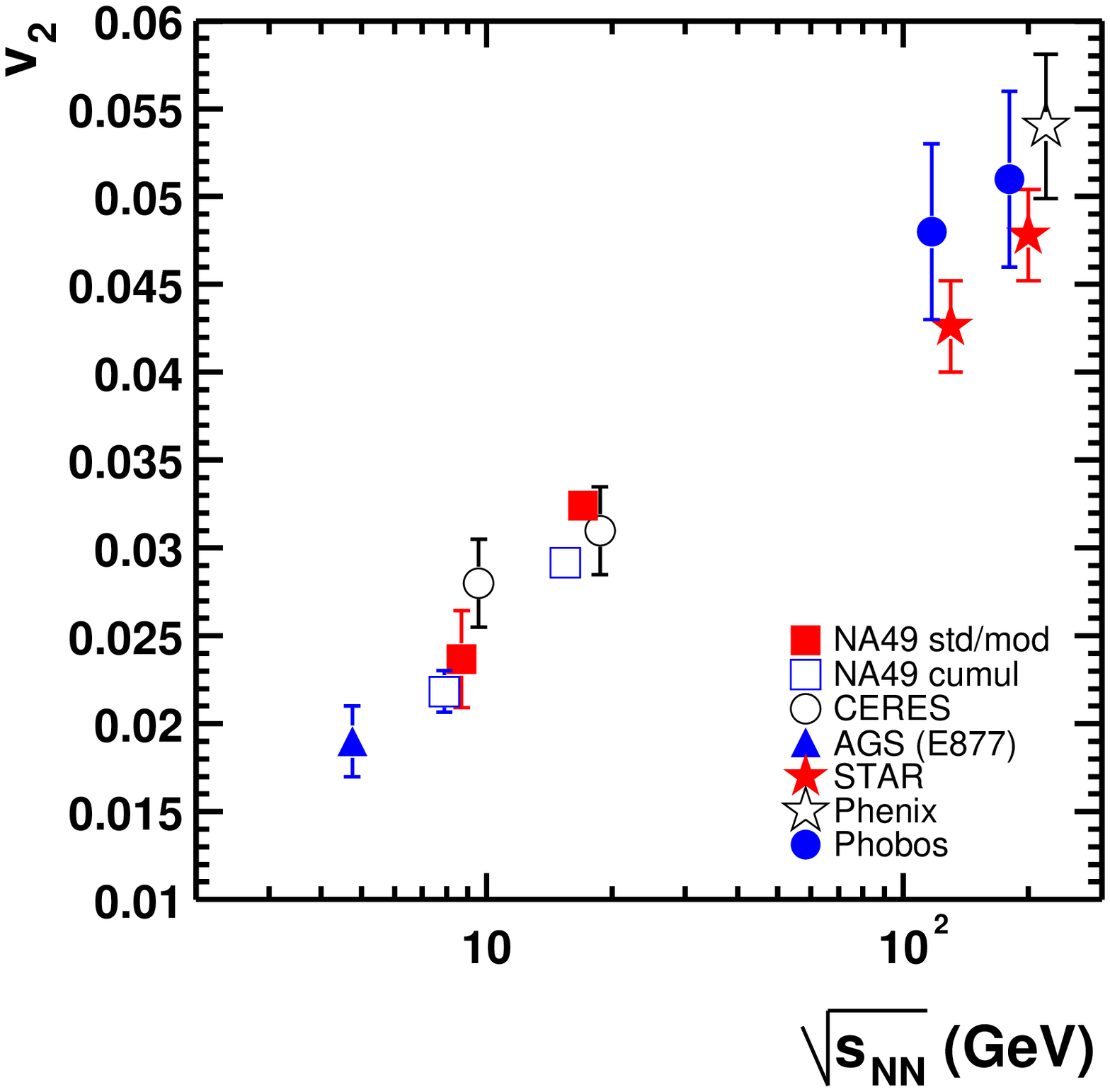}}
\caption{Left panel: elliptic flow as a function of centrality as seen
by STAR (data) compared to hydrodynamic models
(rectangles)~\protect\cite{STAR:v2}. Right panel: peak elliptic flow
as a function of collision energy for ultrarelativistic collisions,
taken from an NA49 compilation~\protect\cite{NA49:v2energy}.}
\label{fig:v2both}
\end{figure}

Non-central heavy ion collisions have an inherent azimuthal
asymmetry. The overlap region of two nuclei is roughly ellipsoidal in
shape. If there is collective motion that develops early in the
collision, this spatial anisotropy can be converted to an azimuthal
asymmetry in the momentum of detected particles. This azimuthal
asymmetry is characterized by a Fourier decomposition of the azimuthal
distribution:
\begin{equation}
dN/d\phi = N_0 (1 + 2 v_1 \cos \phi + 2 v_2 \cos (2 \phi) ),
\end{equation}
where $\phi$ is the azimuthal angle with respect to the reaction
plane\footnote{The true reaction plane is defined by the impact
parameter vector between the gold ions. The experimental results shown
have been corrected for the reaction-plane resolution, which would
otherwise dilute the signal.}.  The left-hand panel of
Fig.~\ref{fig:v2both} shows that the elliptic flow parameter is quite
large, nearly reaching the values predicted by hydrodynamic
models. These models assume a limit of local equilibrium with
collective motion of the bulk ``fluid''.  The right-hand panel of
Fig.~\ref{fig:v2both} shows that this asymmetry is the largest ever
seen at relativistic energies.

\begin{figure}[htbp]
\centerline{
\epsfxsize=63mm\epsfbox{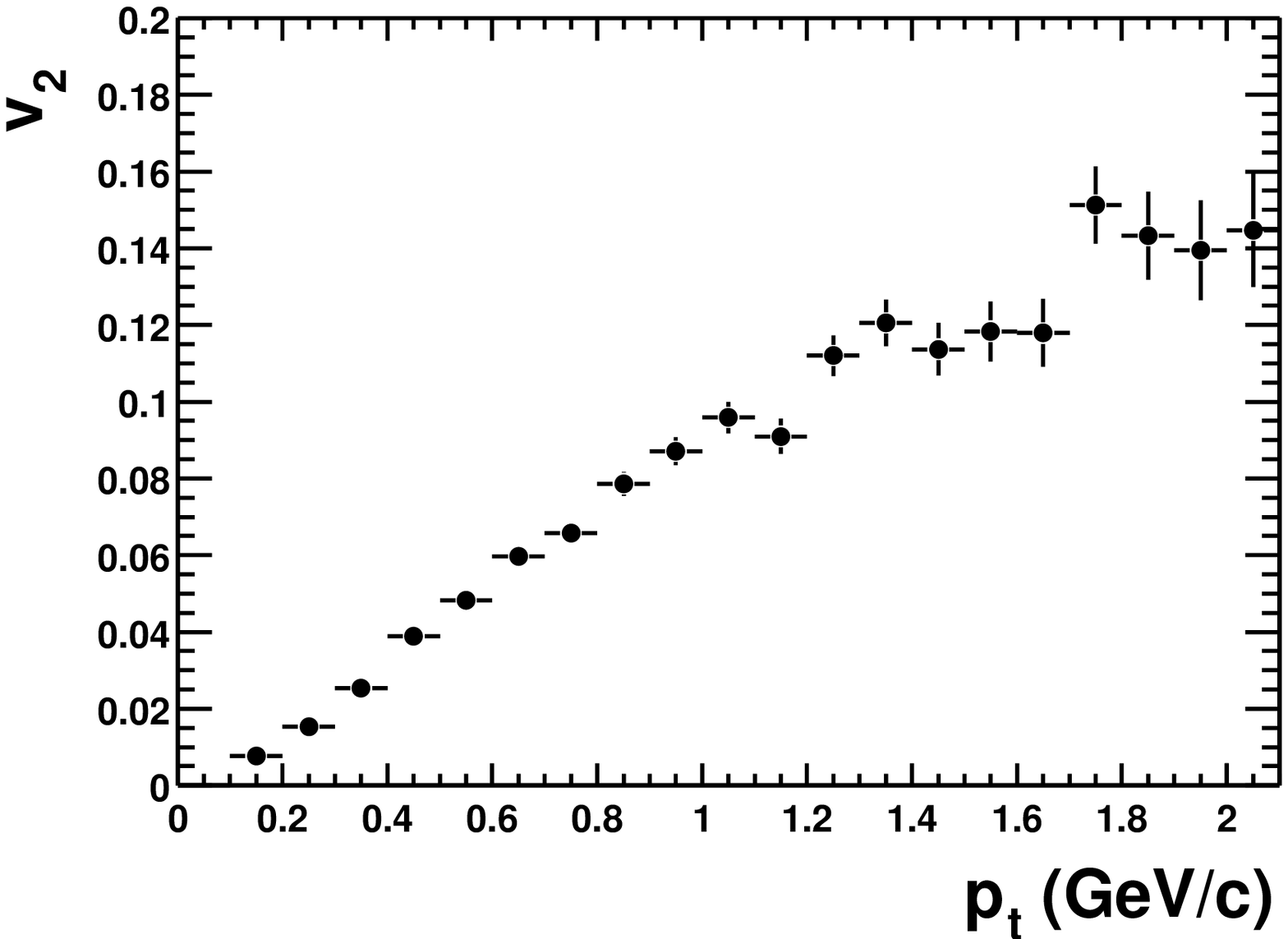}
\epsfxsize=57mm\epsfbox{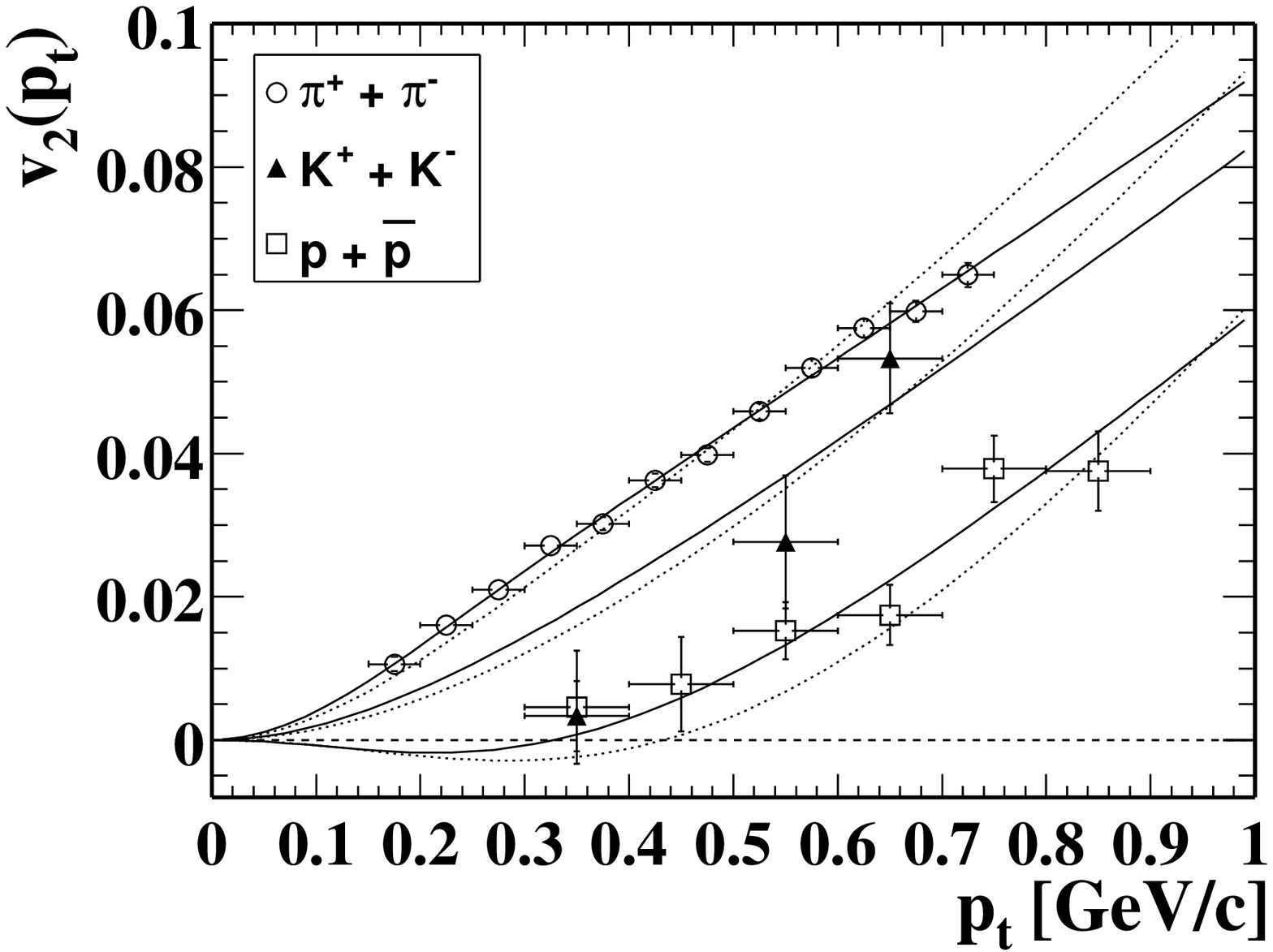}
}
\caption{Elliptic flow versus $\pT$ for all particles (left
panel)~\protect\cite{STAR:v2}, and for identified particles (right
panel) from STAR~\protect\cite{STAR:hydro}. The curves in the right
panel refer to a hydrodynamic model description.}
\label{fig:Starv2Pt}
\end{figure}

Elliptic flow, in addition to indicating that there is collective
motion, can provide information about the type of motion. In
particular, the $\pT$ dependence of elliptic flow can distinguish
between two limits: the low density limit and the hydrodynamic limit
(rapidly expanding opaque source). In the low density limit, some of
the produced particles are absorbed or scattered once (and usually
only once). In this case, for relativistic particles, $v_2$ is nearly
independent of $\pT$. In the hydrodynamic limit, in contrast, we
expect $v_2 \propto \pT$ for moderate values of $\pT$. This effect
comes about because the expansion causes a correlation
between normal space and momentum space, forcing the highest $\pT$
particles to come from the surface, while low $\pT$ particles can come
from anywhere in the volume.  Data from the SPS favor the hydrodynamic
limit~\protect\cite{HL:v2theory}.  The left-hand panel of
Fig.~\ref{fig:Starv2Pt} shows a clear linear relationship between
elliptic flow and transverse momentum at RHIC as well, while the
right-hand panel shows that hydrodynamic models not only describe the
overall trend, but even describe the pions and protons separately.

\begin{figure}[htbp]
\centerline{
\epsfxsize=75mm\epsfbox{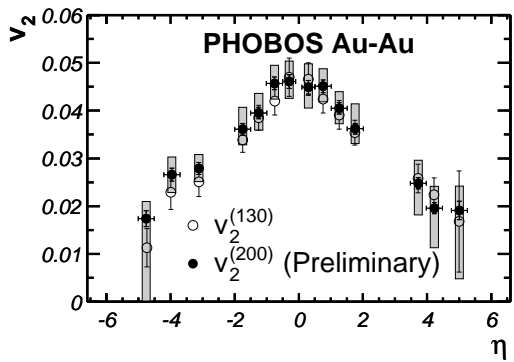}}
\caption{Elliptic flow as a function of pseudorapidity from 
PHOBOS~\protect\cite{PHO:v2}.}
\label{fig:PHOv2}
\end{figure}

Finally, elliptic flow can be examined as a function of
pseudorapidity. The expectation was that the elliptic flow would be
nearly independent of pseudorapidity as the basic physics of RHIC were
expected to be invariant under longitudinal
boosts. Fig.~\ref{fig:PHOv2} shows that $v_2$ is strongly dependent
on pseudorapidity, a result which has still not been explained.

Taken together, these results show clear evidence of collective motion
and suggest a system at or near hydrodynamic equilibrium which is
rapidly expanding in the transverse direction and which does not
exhibit longitudinal boost-invariance.

\subsection{Hanbury-Brown Twiss Effect: More Dynamics}

Intensity interferometery, or the Hanbury-Brown Twiss
effect~\cite{HBT}, is a technique used to measure the size of an
object which is emitting bosons (e.g. photons from a star or pions
from a heavy ion collision). Boson pairs which are close in both
momentum and position are quantum mechanically enhanced relative to
uncorrelated boson pairs. Bosons emitted from a smaller spatial source
are correlated over a broader range in relative momentum, which allows
you to image a static source using momentum correlations.

For a given pair of identical particles, we can define their momentum
difference, $\vec{q}$, and their momentum average, $\vec{k}$. We can
further define the three directions of our coordinate
system~\cite{BP:HBT}:
\begin{itemize}
\item Longitudinal ($R_l$) --- along the beam direction ($\hat{z}$),
\item Outwards ($R_o$) --- In the $(\hat{z},\hat{k})$ plane, $\perp \hat{z}$,
\item Sidewards ($R_s$) --- $\perp \hat{z}$ \& $\perp \hat{k}$.
\end{itemize}

For a boost-invariant source, the measured sidewards radius at low
$\pT$ will correspond to the actual physical transverse (rms) extent
of the source at freezeout, while the outwards radius will contain a
mixture of the spatial and time extent of the source. Particles
emitted earlier look like they are closer to the observer, which
artificially extends the apparent source in the out direction.  In
particular,
\begin{equation}
  R_o^2 - R_s^2 = \beta_\perp^2 \sigma_\tau^2 
- 2 \beta_\perp \sigma_{x\tau} + (\sigma_x^2-\sigma_y^2),
\label{eq:UHfull}
\end{equation}
where $\beta_\perp^2$ is the transverse velocity
associated with $\vec{k}$, $\sigma_\tau$ is the ``duration of
emission'' parameter, $\sigma_x$ and $\sigma_y$ are the geometric size
in the out and side directions, and $\sigma_{x\tau}$ is the space-time
correlation in the out direction. 

In the case of an azimuthally symmetric and transparent source, the last
two terms are taken to be small or zero and we have
\begin{equation}
  R_o^2 - R_s^2 \approx \beta_\perp^2 \sigma_\tau^2.
\label{eq:UHsimple}
\end{equation}

\begin{figure}[htbp]
\centerline{
\epsfxsize=75mm\epsfbox{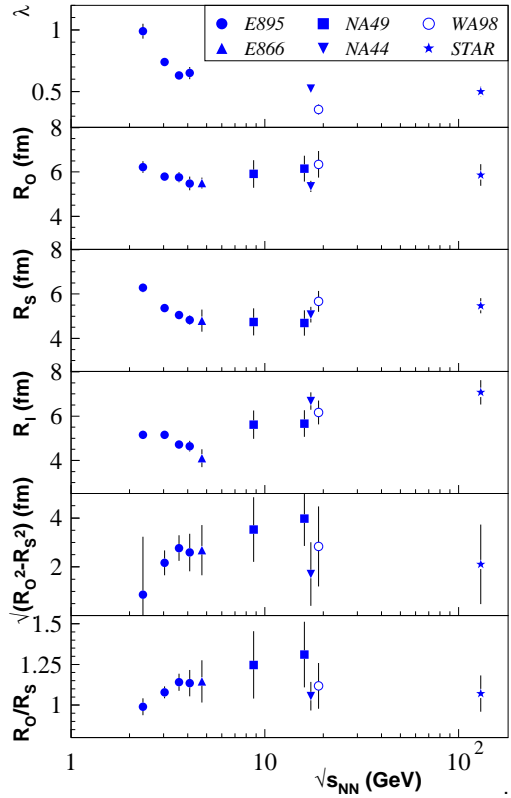}}
\caption{HBT parameters as a function of colliding beam 
energy.~\protect\cite{STAR:HBT}.}
\label{fig:STARhbt}
\end{figure}

Given the assumption of a boost-invariant, azimuthally symmetric and
transparent source, the HBT results from heavy ion collisions have
been perennially confusing. From Eq.~\ref{eq:UHsimple}, we expect
\mbox{$R_o/R_s \ge \sqrt{2}$} since most sources should emit for a
time which is of the same order as their size. Some models of the
Quark-Gluon Plasma predict an even larger value for this ratio as the
plasma might need to emit particles over a long time duration in order
to get rid of the entropy~\cite{Dirklog}. However, as can be seen in
Fig.~\ref{fig:STARhbt}, $R_o/R_s$ is basically unity at RHIC energies,
na\"{\i}vely implying an instantaneous emission of particles over a
moderately large volume.

This situation, along with the modest values of $R_l$,
has been termed the ``HBT puzzle''. Primarily, though, these data
indicate a need to improve the modeling of the collision. If you
consider a source which is opaque, rapidly expanding and also not
boost invariant, the meaning of $R^2_o-R^2_s$ changes since we must
use Eq.~\ref{eq:UHfull} and not Eq.~\ref{eq:UHsimple}.  Opacity
reduces the apparent $R_o$ value since you only see the part of the
source closest to you in the out direction. Transverse expansion along
with opacity will decrease the ratio further since particles emitted
later are also emitted closer to the viewer, reducing the magnitude of
$R_o$. Finally, a general longitudinal expansion (not just coasting)
must be taken into account since we know that the source is not boost
invariant. This effect would explain the small size of $R_l$ and has
also been shown~\cite{Dirk} to reduce the ratio $R_o/R_s$.

So, while HBT and elliptic flow have not been successfully described
in full detail by the hydrodynamic models yet, the qualitative message
they provide is very similar. The source is rapidly expanding
(probably in all three dimensions), opaque, and can be described as
``hydrodynamically equilibrated bulk matter''.

\subsection{Characterizing the Bulk Matter}

Having established that the system has a large number of particles as
well as collective behavior, we can now proceed to consider bulk
quantities such as the temperature and baryon chemical potential of
the system.

In conventional, static, thermodynamic systems, the temperature can be
measured by directly measuring the average energy per particle. In a
very dynamic system, such as a heavy ion collision, we have to
separate the energy contributed by collective motion from the thermal
energy.  To do this, we make use of the fact that the collective velocity
contributes more to the momentum of heavy particles than to lighter
particles. Thermal fits~\cite{thermfit} to $\langle \pT \rangle (m)$
yield a temperature of approximately 100~MeV and an average transverse
expansion velocity of 0.55~c. This large expansion velocity supports
the picture given by the elliptic flow and HBT.

\begin{figure}[htbp]
\centerline{
\epsfxsize=11cm\epsfbox{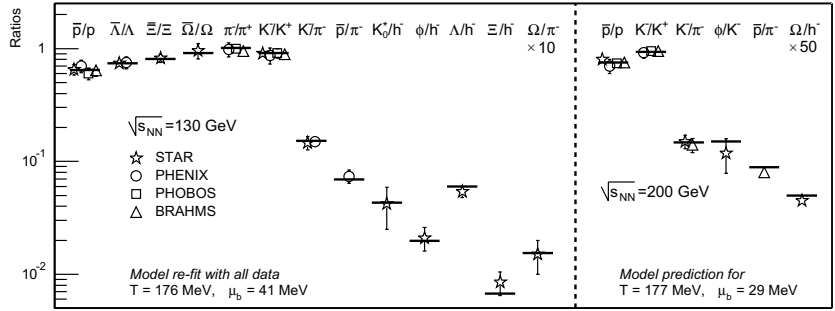}}
\caption{Particle abundance ratios and a thermal fit for the two
highest RHIC energies~\protect\cite{TU:QM2002}.}
\label{fig:Ullrich02}
\end{figure}

Another thermometer is provided by the fact that the ratios of
particles of different masses are sensitive to the temperature. In
addition, ratios of particles with the same mass, but different quark
content, such as $\bar{p}/p$ and $K^-/K^+$, are sensitive to the
balance between matter and antimatter, characterized by the baryon
chemical potential $\mu_B$.  Positive values of $\mu_B$ refer to a
matter (baryon) excess in a system. Fig.~\ref{fig:Ullrich02} shows
particle abundance ratios and a thermal fit. This fit yields a
constant temperature of 176--177~MeV at both energies, but a falling
baryon density (41~MeV at 130~GeV and 29~MeV at 200~GeV). The falling
baryon density is expected. Higher energy collisions dilute the fixed
initial baryon excess from the original gold nuclei and also make it harder
to transport the baryon excess to midrapidity.

We are immediately faced with a dilemma: our kinetic thermometer,
based on energy per particle, indicated a temperature of $\sim
100$~MeV, while our chemical thermometer, based on particle
abundances, indicated a much higher temperature of $\sim 175$~MeV. The
resolution of this paradox lies in the fact that only inelastic
collisions can change the particle abundances while both elastic and
inelastic collision serve to equalize the energy between
particles. Using the terminology of cosmology, we can define an
approximate ``freezeout hypersurface'' which contains the spacetime
points of the final collisions suffered by each particle.  In the case
of a heavy ion collision, the chemical freezeout can occur earlier
than the kinetic freezeout. This resolves our dilemma, but with the
unavoidable consequence of making our picture of the collision
somewhat more complicated. It should be noted that the HBT results are
actually imaging the kinetic freezeout boundary as the source.

\subsection{``Little Bang Cosmology''}

As in cosmology, we are interested in understanding what happened
before the freezeout. We can estimate the energy density from the
transverse energy produced in the collision and the cylindrical volume
occupied shortly after the collision occurred. This leads to the 
formula~\cite{Bj:energy}:
\begin{equation}
\epsilon = \frac{1}{\pi R^2} \frac {1}{c \tau_0} \frac{dE_T}{dy}
\label{eq:Bj}
\end{equation}
where the radius R is the nuclear radius and $\tau_0$ is the time it
takes for the transverse energy to be effectively equilibrated 
(0.2--1.0~fm/c).

\begin{figure}[htbp]
\centerline{
\epsfxsize=10cm\epsfbox{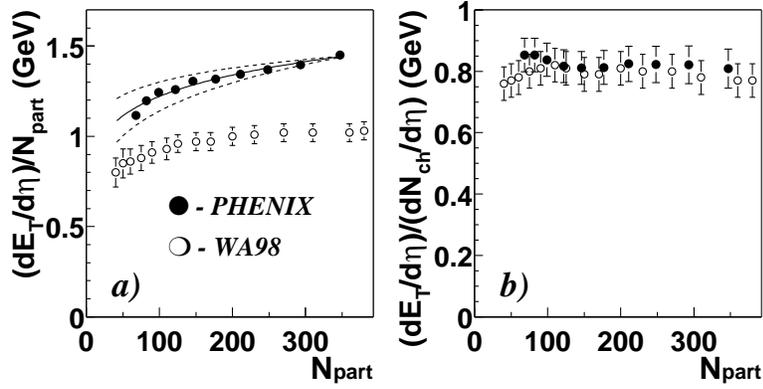}}
\caption{Transverse energy at midrapidity as a function of centrality
for 130~GeV and 17~GeV collisions. Left panel: per participating
nucleon, right panel: per produced particle. Data taken
from~\protect\cite{PHX:Et}.}
\label{fig:PhxEt}
\end{figure}
\begin{figure}[htbp]
\centerline{
\epsfxsize=10cm\epsfbox{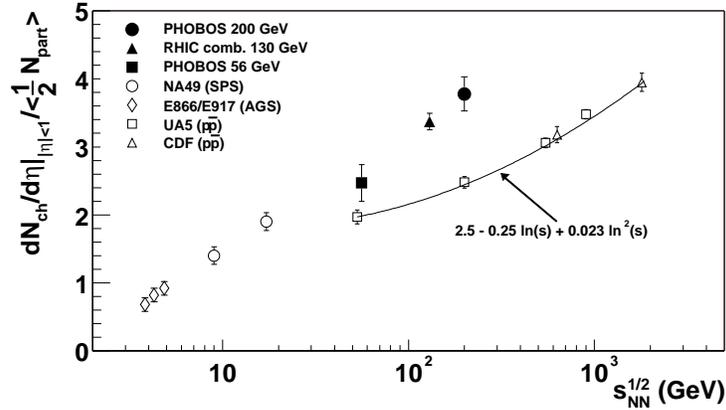}}
\caption{Charged particle multiplicity per participating nucleon pair
at midrapidity as a function of beam energy~\protect\cite{PHO:mult}.}
\label{fig:PHOmult}
\end{figure}
\begin{figure}[htbp]
\centerline{
\epsfxsize=8cm\epsfbox{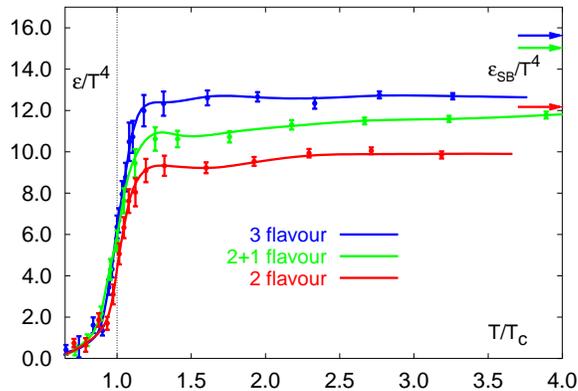}}
\caption{Energy density scaled by $T^4$ (natural units $\hbar=c=k=1$)
as as a function of temperature scaled by the critical
temperature~($T/T_c$). The arrows on the right indicate the
Stefan-Boltzmann values for an ideal non-interacting gas. Figure taken
from Ref.~\protect\cite{Karsch}.}
\label{fig:lattice}
\end{figure}

Figure~\ref{fig:PhxEt} (right panel) shows that the transverse energy
per particle is about 800~MeV at RHIC while Fig.~\ref{fig:PHOmult}
shows the multiplicity.  Combining these results, using Eq.~\ref{eq:Bj}
yields $\epsilon=$5--25 ${\rm GeV/fm}^3$ for central
collisions at the highest RHIC energy. Fig.~\ref{fig:lattice} shows
the theoretical relationship, based on lattice QCD
calculations~\cite{Karsch}, between energy density and
temperature. The expected $T^4$ dependence of an ideal gas has been
divided out, leading to a a constant value for high temperature,
proportional to the number of degrees of freedom in the quark-gluon
plasma phase.

Combining the data with the theoretical curves leads to an estimated
initial temperature of $300 \pm 50$~MeV for central \AuAu\ collisions
at the top RHIC energy. This is significantly higher than the
theoretical transition temperature of $\sim 170$~MeV. A similar
exercise at the top CERN-SPS energy $\sqrtsnn=$17~GeV, yields an
estimated initial temperature of $240 \pm 50$~MeV. It should be noted
that if we assume a hadronic description rather than a phase
transition, the number of degrees of freedom should actually be lower,
implying an even higher initial temperature (about twice as high).
This means that the estimated initial temperatures of $\sim 300$ and
240 MeV for RHIC and CERN actually represent lower limits.

\subsection{Summary: Bulk Matter}

\begin{figure}[htbp]
\vspace{15mm}
\hspace{1cm} 
\epsfxsize=7cm\epsfbox{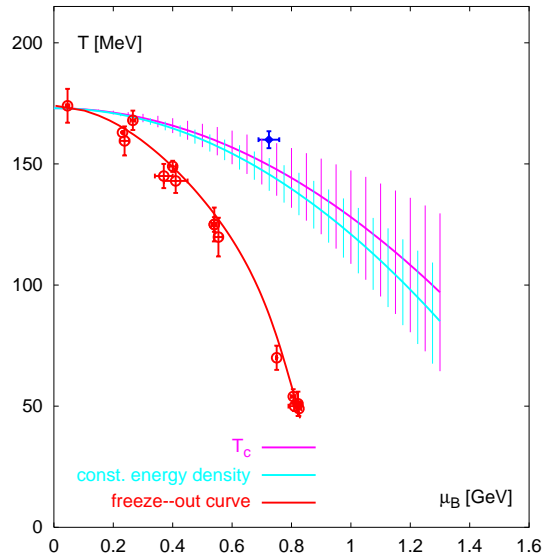}
\vspace{-5mm}
\caption{Phase diagram of heavy ion collisions from
Ref.~\protect\cite{phasediag}. The data points represent heavy ion
collisions over a broad range of energies. The curve through the data
points represents a fixed energy per particle. The upper band represents
an estimate of the phase boundary. The lower band represents a
constant energy density (0.6~${\rm GeV/fm^3}$). The isolated point
above the theory curves represents a theoretical critical point.}
\label{fig:phase}
\end{figure}

Figure~\ref{fig:phase} shows the phase diagram based on the chemical
freezeout points measured at various energies in heavy ion collisions
including the 130~GeV point from RHIC. The 200~GeV point from RHIC
would be at basically the same temperature, but $\mu_B=29$~MeV rather
than $\mu_B=41$~MeV. The curve through the data implies freezeout at a
fixed energy per particle of about 1 GeV, while the bands indicate the
theoretical expectation for the transition between confined and
deconfined matter. The initial temperatures estimated for both RHIC
and CERN are not shown, but they would lie above the theory curve,
with the RHIC temperature being $300 \pm 50$~MeV.  The constant freezeout
temperature for high energy ion collisions, appearing at the
theoretical boundary between confined and deconfined matter is
provocative. It could be an accident, but it is similar to a situation
where you have a detector which only detects liquid, you determine
indirectly that you created matter at $200^\circ {\rm C}$, and you
directly detect droplets of water at a temperature of $100^\circ {\rm
C}$.

To summarize this section, we have produced a dense, hot, rapidly
expanding bulk matter state. We have seen a universal freezeout
curve and it is suggestively close to the expected boundary between
deconfined and confined matter. Furthermore, we have indications that
the initial collision reaches energy densities (and therefore
temperatures) well in excess of that expected to be needed for
deconfinement.

Efforts to probe this state quantitatively are just beginning, but
show promise. This will be the subject of the next section.

\section{Probing the Earliest, Hottest Part of the Collision}

While the freezeout temperature measurements are on solid footing, the
estimates presented above for the initial temperature are indirect.
As the RHIC program develops, we can go beyond these qualitative
discussions of the early times and start probing them more
quantitatively.

\subsection{Electromagnetic and Hidden Flavor Probes}

Perhaps the cleanest method, from a theoretical perspective, would
be to examine thermal photons and leptons that originate from the
early part of the collision when the temperature was higher. These
weakly interacting particles are expected to decouple thermally (or
``freezeout'') from the bulk strongly interacting matter much earlier
than hadrons. Combined with a measurement of the energy density this
would effectively measure the number of degrees of freedom in the
initial state. While it is theoretically very clean, this measurement
is experimentally very challenging. A typical central collision at
RHIC produces thousands of neutral pions which decay into thousands of
photons in each event and serve as a background to this measurement.
A typical RHIC detector also has literally tons of material in which
background photons (and leptons) can be produced.

Despite the difficulty, these measurements and fits have been
attempted at the SPS, both in terms of direct photon
spectra~\cite{WA98:photons} and thermal
leptons~\cite{NA50:muons}. These results lead to an estimated initial
temperature at the SPS of $\sim 200$~MeV, consistent with our estimate
above for partonic matter.  These results, however, are very sensitive
to details of how the backgrounds are handled.

\begin{figure}[htbp]
\centerline{
\epsfxsize=7cm\epsfbox{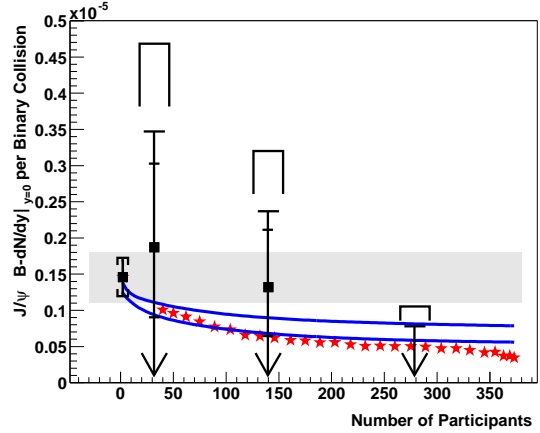}}
\caption{Measured yield of J/$\psi$ by NA50 at CERN (stars) and by
PHENIX at RHIC (black squares and arrows)~\protect\cite{Phx:Jpsi}.}
\label{fig:jpsi}
\end{figure}

Hidden heavy flavor measurements (strangeness, charm, and beauty) also
show promise as potentially sensitive probes of the density of the
medium and of chiral symmetry restoration.  Fig.~\ref{fig:jpsi}
shows the suppression of the J/$\psi$ (hidden charm) compared to
collision scaling at RHIC and the SPS. Sensitivity can also be
found in the mass, line shape, and yield of the $\phi$ particle
(hidden strangeness), seen by its hadronic and leptonic decay
modes~\cite{Phx:phis}. So far at RHIC, these measurements suffer
from lack of statistical power.

One common denominator that would make many of these signals clearer
would be a clean measurement of open heavy flavor (D and B
particles). These measurements should be forthcoming from RHIC
following upgrades to the detectors and improved luminosity from the
collider.

\subsection{Hadron Suppression: Jet Quenching?}

In addition to measuring the initial temperature, we would
like to have a more direct measure of the energy density of the bulk
matter that we have created. One handle on this quantity is to study
the behavior of high momentum particles in heavy ion collisions. In
particular, partons with relatively high transverse momentum are
predicted to lose energy when traveling through dense matter, in a
phenomenon known as ``jet quenching''. The amount of energy loss is
proportional to the energy density of the matter traversed, so this is
potentially a very sensitive probe.

All four experiments at RHIC measured particle
spectra~\cite{STAR:Raa130,PHO:Raa200,Phx:Raa200,Star:Raa200,Bra:Raa200,Phx:FlagPRL}.
These spectra need to be compared to a reference sample, appropriately
scaled. The simplest such reference sample is to consider each NN
collision in the initial AA collision geometry as being
independent. This leads us to define a ``nuclear modification factor'':
\begin{equation}
R_{AA}(p_T) \equiv
\frac{\sigma_{inel}^{pp}}{N_{coll}}
\frac{d^2N^{AA}/dp_Td\eta}{d^2\sigma^{pp}/dp_Td\eta}.
\label{eq:RAA}
\end{equation}
At high momentum, ($p_T> 2$~GeV/c) this ratio should approach unity if
the collisions are independent and the jets are not affected by the
material. Jet energy loss in the medium should show up as a
suppression of high momentum hadrons. In lower energy AA
collisions~\protect\cite{SPSCronin} and \pA\ collisions~\cite{Cronin},
an excess has been observed rather than a suppression.  This effect is
interpreted as being caused by multiple scattering during the initial
collision.

\begin{figure}[htbp]
\centerline{
\epsfxsize=85mm\epsfbox{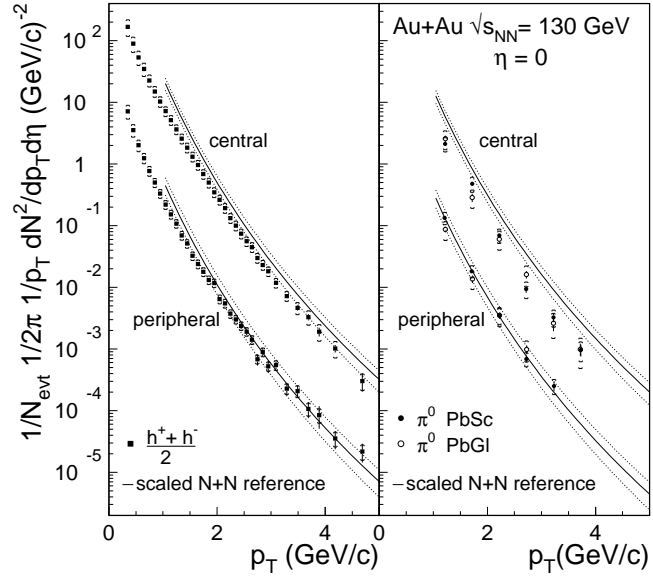}}
\caption{Yields from PHENIX for a) charged particles and b) neutral pions for
central and peripheral 130~GeV \AuAu\ collisions compared to scaled reference
samples~\protect\cite{Phx:FlagPRL}.}
\label{fig:Phenixraw}
\end{figure}

\begin{figure}[htbp]
\centerline{
\epsfxsize=75mm\epsfbox{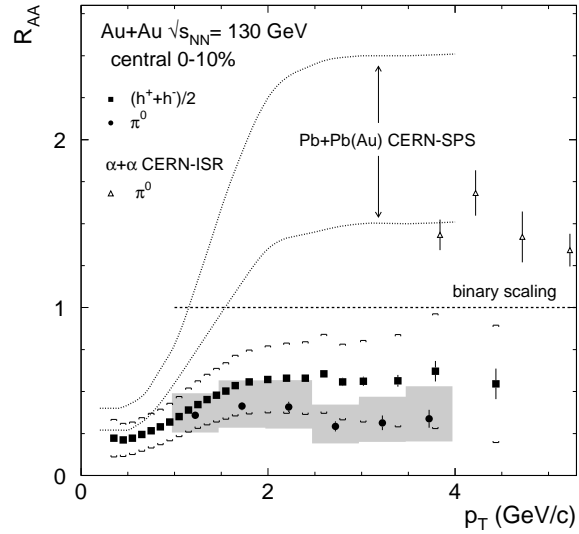}}
\caption{Nuclear modification factor for charged particles
and neutral pions for 17~GeV \PbPb\ collisions at CERN and central
130~GeV \AuAu\ collisions from PHENIX~\protect\cite{Phx:FlagPRL}.}
\label{fig:PhxRaa}
\end{figure}

\begin{figure}[htbp]
\centerline{
\epsfxsize=80mm\epsfbox{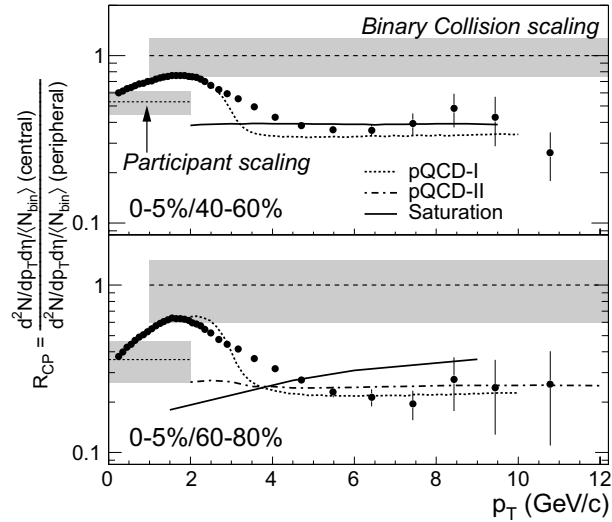}}
\caption{Scaled ratio of charged particle production in central to
peripheral \AuAu\ collisions at 200~GeV from STAR, using two different
choices of peripheral data~\protect\cite{Star:Raa200}.}
\label{fig:StarRaa}
\end{figure}

The results at RHIC energies are strikingly different from lower
energy data as can be seen in Fig.~\ref{fig:Phenixraw} from
PHENIX. Invariant yields for produced particles in central and
peripheral 130~GeV \AuAu\ data are compared to a scaled-up pp
reference sample. For $\pT>2$~GeV, the peripheral data scales as
expected, while the central \AuAu\ data shows a substantial
suppression. The dramatic difference between the different energies is
even more apparent in Fig.~\ref{fig:PhxRaa} where the scaled reference
data are divided out to yield $R_{AA}$, the nuclear modification
factor of Eq.~\ref{eq:RAA}. Clearly, something qualitatively different
is occurring at RHIC energies. Similar results were seen by STAR at
130 GeV~\cite{STAR:Raa130} and all four experiments at 200
GeV~\cite{PHO:Raa200,Phx:Raa200,Star:Raa200,Bra:Raa200}. Since the
peripheral data scales as expected, it is also possible to measure
hadron suppression by taking the ratio of central/peripheral data,
scaled by the ratio of $N_{coll}$. At 200~GeV, this technique was used
to establish that this hadron suppression persists to very high
transverse momentum, as seen in Fig.~\ref{fig:StarRaa}.  As indicated
above, this hadron suppression may be a signature of jet quenching, in
which case we have clear evidence of a system with very high energy
density.

\begin{figure}[htbp]
\centerline{
\epsfxsize=49mm\epsfbox{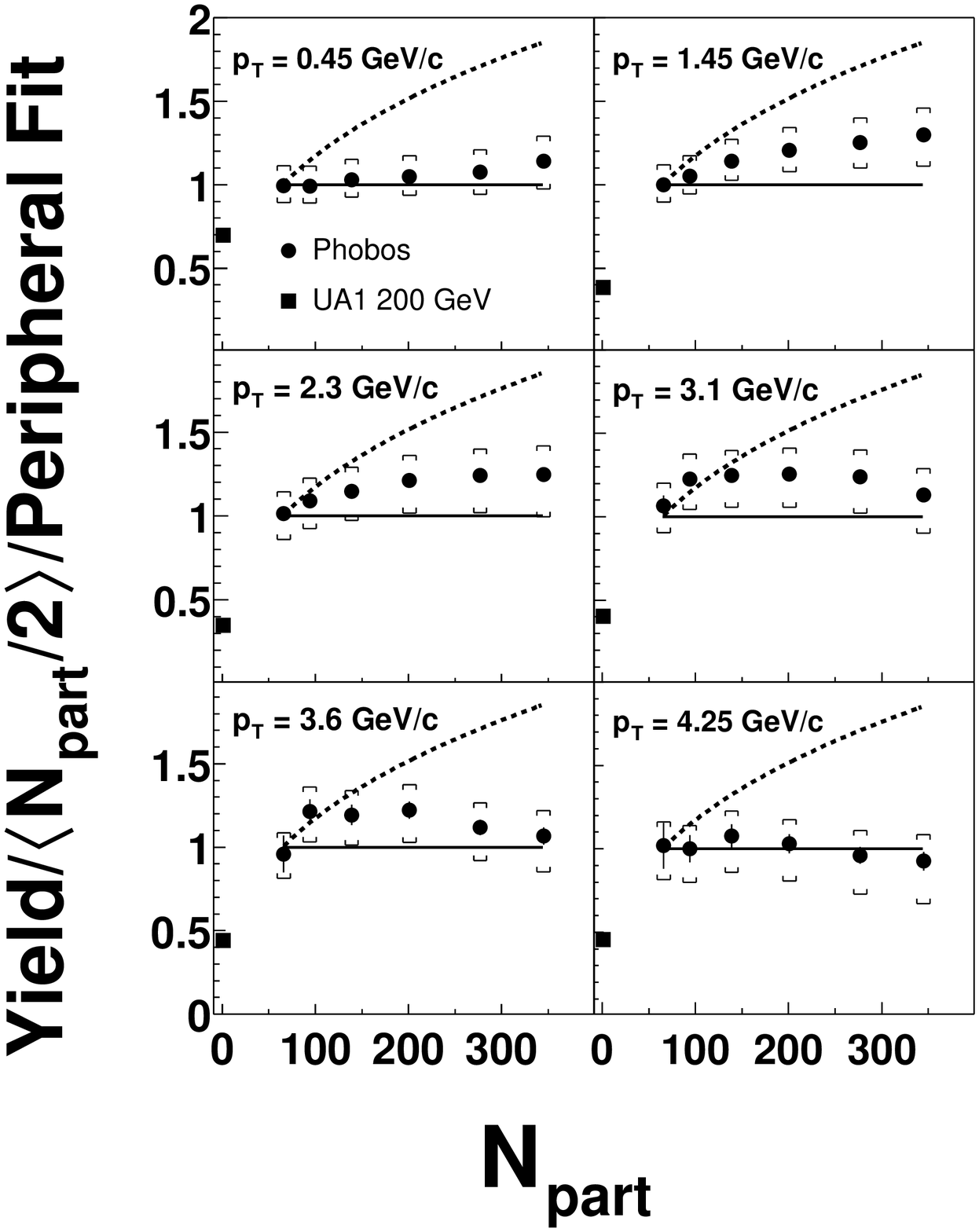}
\epsfxsize=64mm\epsfbox{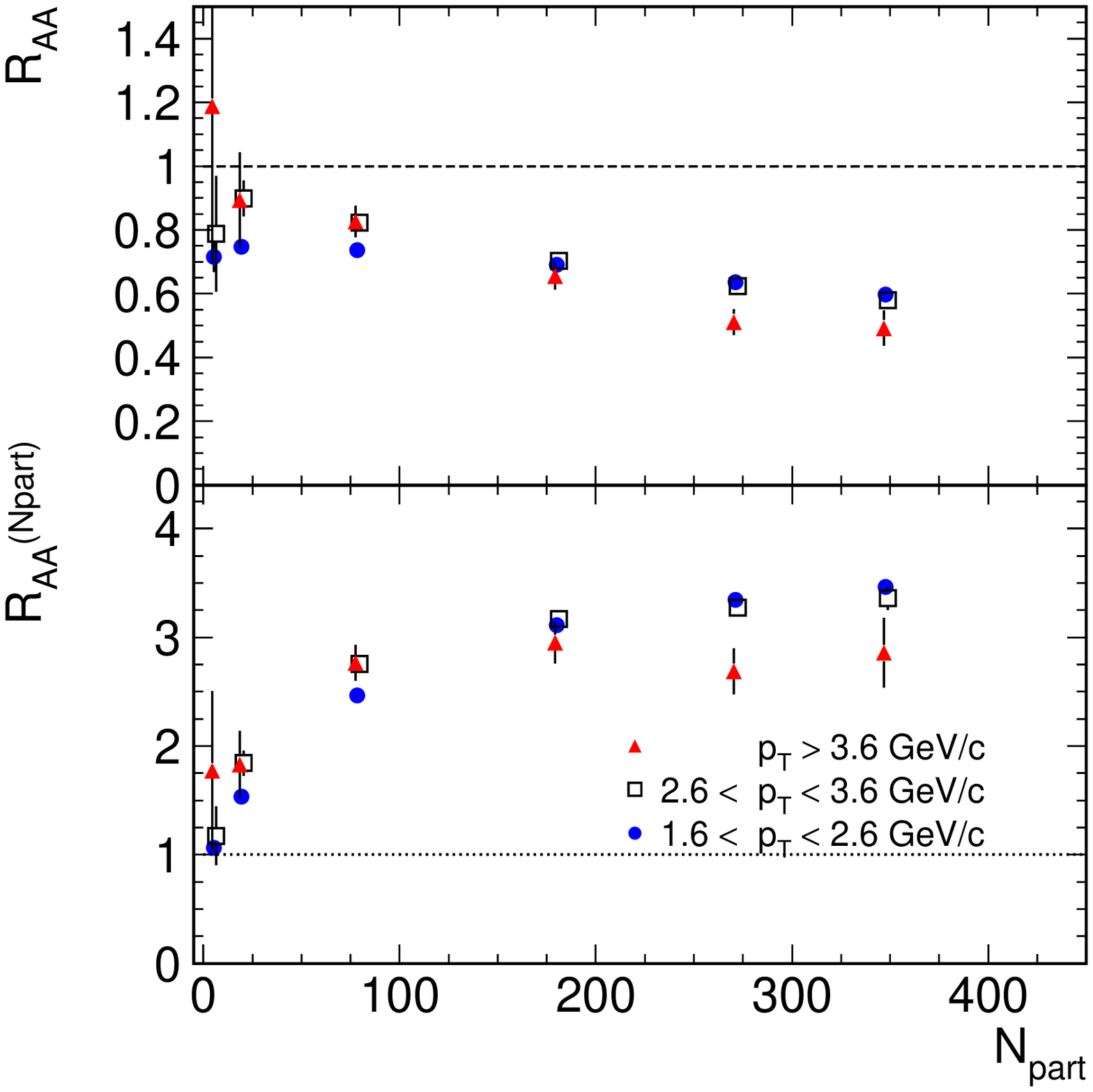}
}
\caption{Scaling of charged hadron yields from \AuAu\ collisions from
PHOBOS and PHENIX. Left panel: Yield scaled per participant pair
($\avgNp/2$) normalized to the yield of the most peripheral bin
(45--50\%)~\protect\cite{PHO:Raa200}. Right panel (lower plot): Yield
scaled per participant pair with reference to \pp\
data~\protect\cite{PhxNpart}.  }
\label{fig:Npartscale}
\end{figure}

Another view of this hadron suppression, from PHOBOS and PHENIX, shows
how strong the effect is.  Fig.~\ref{fig:Npartscale} (left
panel) shows the yield in \AuAu\ collisions for fixed values of $\pT$,
scaled by mid-central data ($\avgNp \sim 65$) and normalized by
$\avgNp$.  Consider the lower-right hand plot, with $\pT=4.25$~GeV/c.
The dashed curve shows the expectation if $N_{coll}$ ($A^{4/3}$)
scaling held true over the centrality range
shown, while the solid line shows the expectation of $N_{part}$
($A^1$) scaling .  For $N_{part}>65$, we see approximate $A^1$ scaling
of high $\pT$ particle production. Fig.~\ref{fig:Npartscale} (right
panel) shows a similar result, normalized to \pp\ data, from
PHENIX. The lower plot is normalized per participant and the result
for $\pT>3.6$~GeV/c is relatively flat for $\avgNp>80$.  This particular
form of high $\pT$ suppression could be an indication that jet
quenching reaches a geometric maximum involving one power of length
scale $R_{Au} \propto N_{part}^{1/3}$ (see e.g. Ref.~\cite{bmquench}).

Another piece of evidence in favor of the jet quenching interpretation
for this data comes from STAR.  Jets in pp collisions can be seen by
triggering on a high momentum particle and then looking for
correlations of moderate $\pT$ particles azimuthally. In pp
collisions, this leads to a clear two-jet signal with a cluster of
particles near the trigger particle in azimuth and another cluster at
$\Delta \phi=\pi$ (back-to-back correlation). This signal indicates
that jets are created and acquire large transverse momentum in
conventional $2\rightarrow 2$ parton scattering processes and that the
jets survive. For peripheral \AuAu\ collisions, one expects a similar
result as found in pp, with a small correction due to correlations
induced by elliptic flow\footnote{Since particles are preferentially
produced in the event plane, a trigger particle in the event plane
will tend to pick up particles at $\Delta \phi = 0$ or $\pi$.  This
means that the appropriate reference is $C_2(p+p) + A (1+2v_2^2
\cos(2\Delta\phi)).$}.

\begin{figure}[htbp]
\centerline{
\epsfxsize=50mm\epsfbox{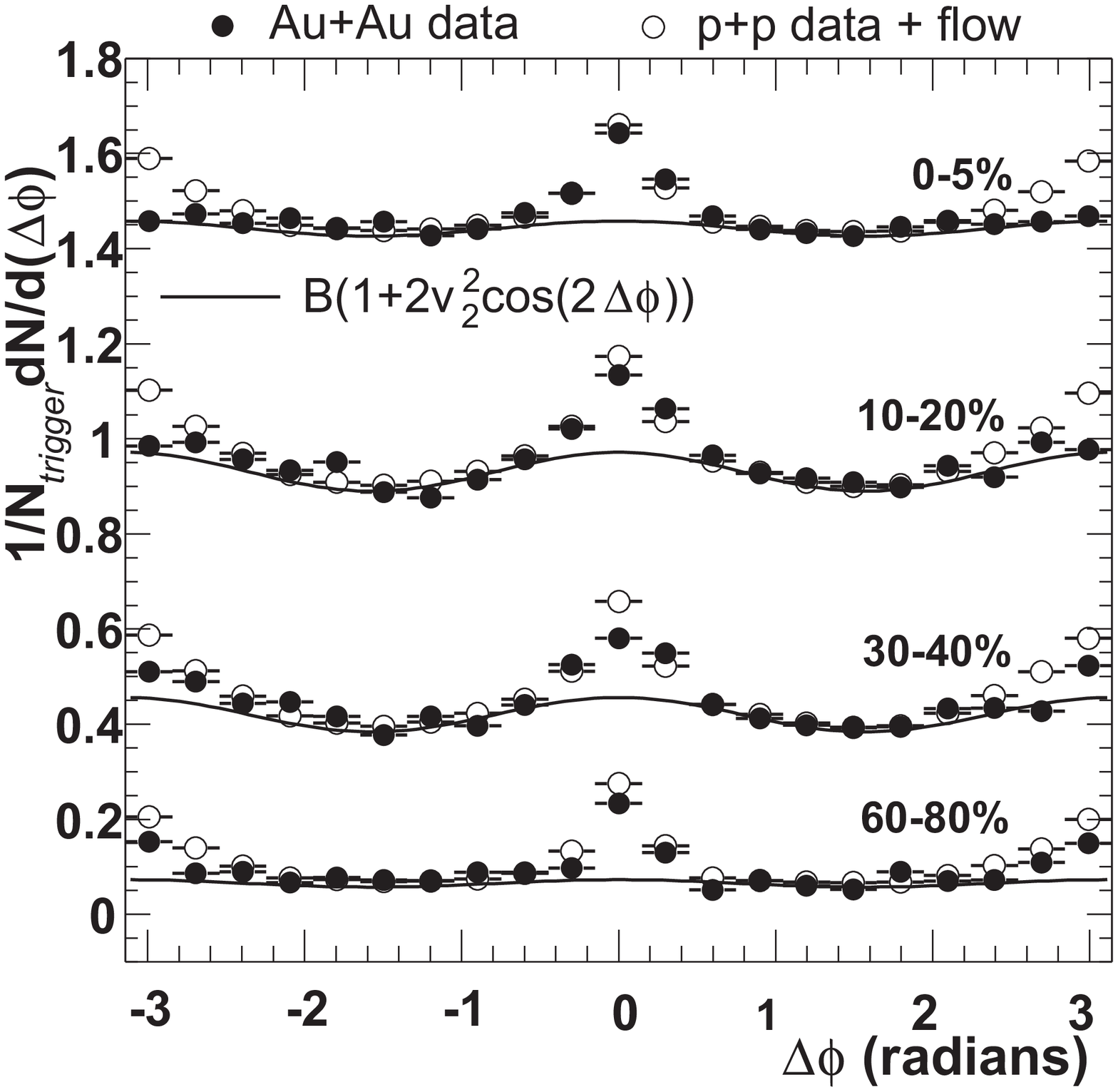} 
\epsfxsize=66mm\epsfbox{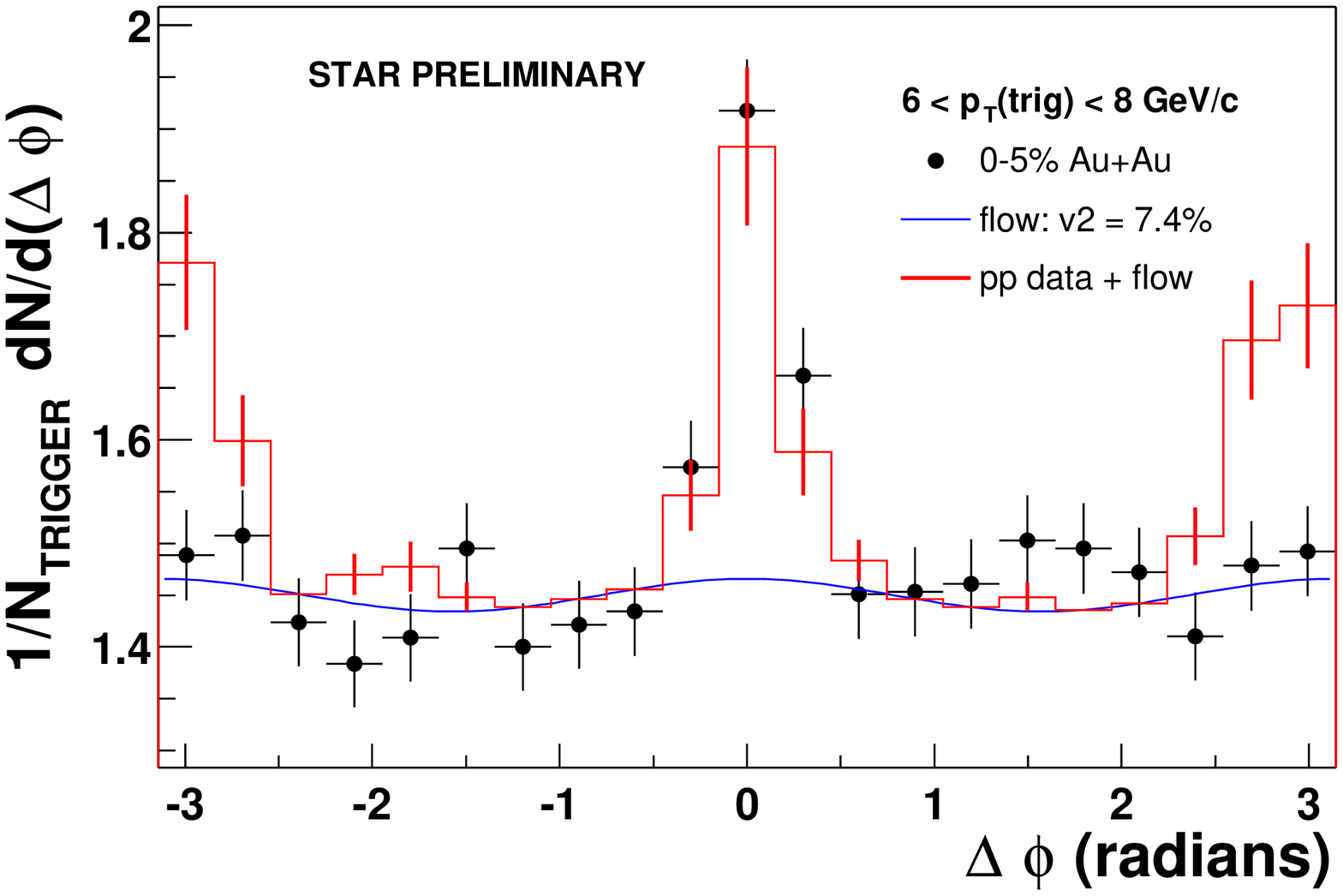}
}
\caption{Azimuthal correlation functions from
STAR~\protect\cite{STARbb}.  Left panel: Data from peripheral \AuAu\
collisions (filled circles) compared to a jet+flow reference sample
(open circles) and a flow-only reference curve. Right panel: Data from
central \AuAu\ collisions (points) compared to a jet+flow reference
sample (upper curve) and a flow-only reference curve (lower curve).}
\label{fig:STARbb}
\end{figure}

Figure~\ref{fig:STARbb} (left panel) shows that the reference sample
constructed from \pp\ collisions and the measured elliptic flow
successfully describes the peripheral \AuAu\ data:
jets are created back-to-back and survive.  In contrast,
Fig.~\ref{fig:STARbb} (right panel) shows the result for central
\AuAu\ data. In this case, the azimuthal correlation function agrees
with the jet+flow reference for $\Delta\phi\sim 0$ while it agrees
with the flow-only reference for $\Delta\phi\sim \pi$.  This means
that the near-side jet survives, but the away-side jet disappears. The
main point here is that this measurement shows that the hadron
suppression is a jet phenomenon. If back-to-back jets are indeed
produced as expected in central \AuAu\ collisions, then the away-side
jet is quenched by the bulk matter.

\subsection{Is Jet Quenching the Only Possible Explanation?}

Triggered by the observation that the scaling is approximately
proportional to $N_{part}$, or $A^1$, at large $\pT$, Kharzeev, Levin,
and McLerran showed that the ``suppression'' of jets in \AuAu\
compared to NN could simply be due to initial state effects
already present in the gold nuclear wavefunction~\cite{KLM}. 
Parton recombination (or saturation) can cause gluons from
different nucleons in the gold nuclei to recombine, leading to a
smaller number of partons with a higher transverse momentum per
parton. Qualitatively, this is difficult to distinguish from jet
quenching since it reproduces both effects:
\begin{enumerate}
\item There are fewer high $\pT$ jets than expected because the gold
      nuclei are not simple linear superpositions of nucleons and
      there are just fewer quarks and gluons to begin with than expected.
\item Jets do not necessarily come out back-to-back.
      The usual argument for back-to-back jets assumes two incoming
      partons with $\pT \sim 0$ followed by a large angle $2
      \rightarrow 2$ scatter into two back-to-back jets. However,
      multiple parton collisions in the initial state lead to 
      partons with non-zero $\pT$ compensated by multiple
      partners, which need not appear at $\Delta\phi=\pi$.
\end{enumerate}

\begin{figure}[htbp]
\centerline{
\epsfxsize=75mm\epsfbox{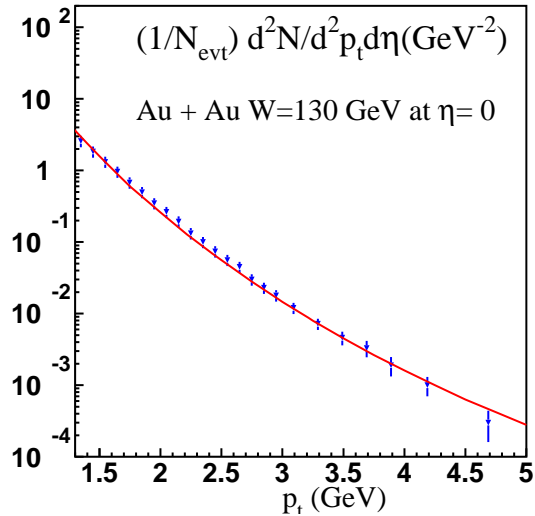}}
\caption{Invariant yield from PHENIX compared to the KLM saturation
         model~\protect\cite{KLM}.}
\label{fig:KLMphx}
\end{figure}

These authors also showed that the initial state saturation model
could be made to agree quantitatively with the data~(see
Fig.~\ref{fig:KLMphx}), including the effect of approximate $N_{part}$
scaling~\cite{KLM}. 

The saturation model also describes the overall particle
production. In fact, this ability is more natural since parton
saturation effects are strongest for the softest partons where the
parton densities are the highest. The saturation model relates the
gluon distribution at low x in deep inelastic scattering from
protons with the energy, centrality, and pseudorapidity dependence of
particle production in heavy ion collisions~\cite{KL130}.

\begin{figure}[htbp]
\centerline{
\epsfxsize=45mm\epsfbox{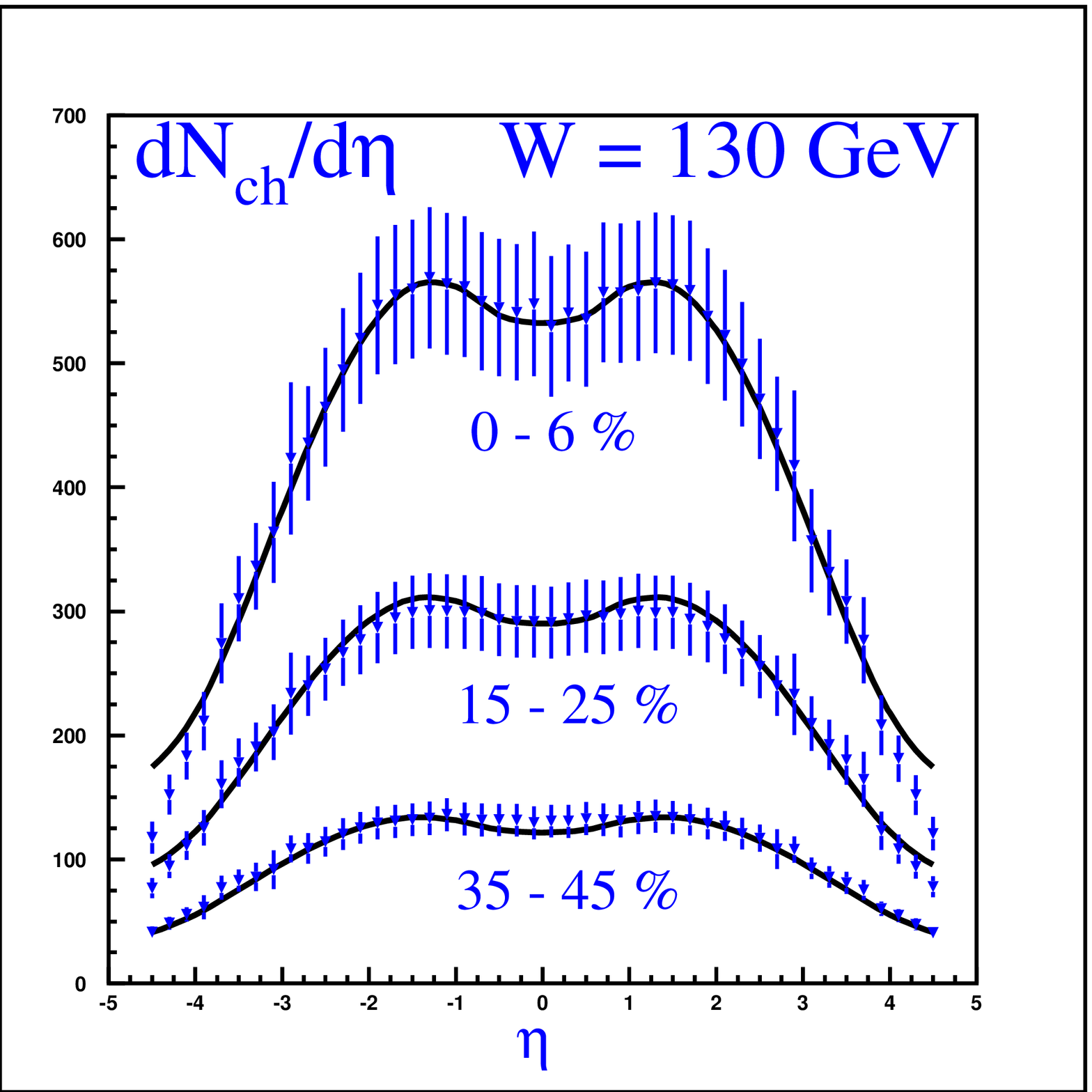}
\epsfxsize=70mm\epsfbox{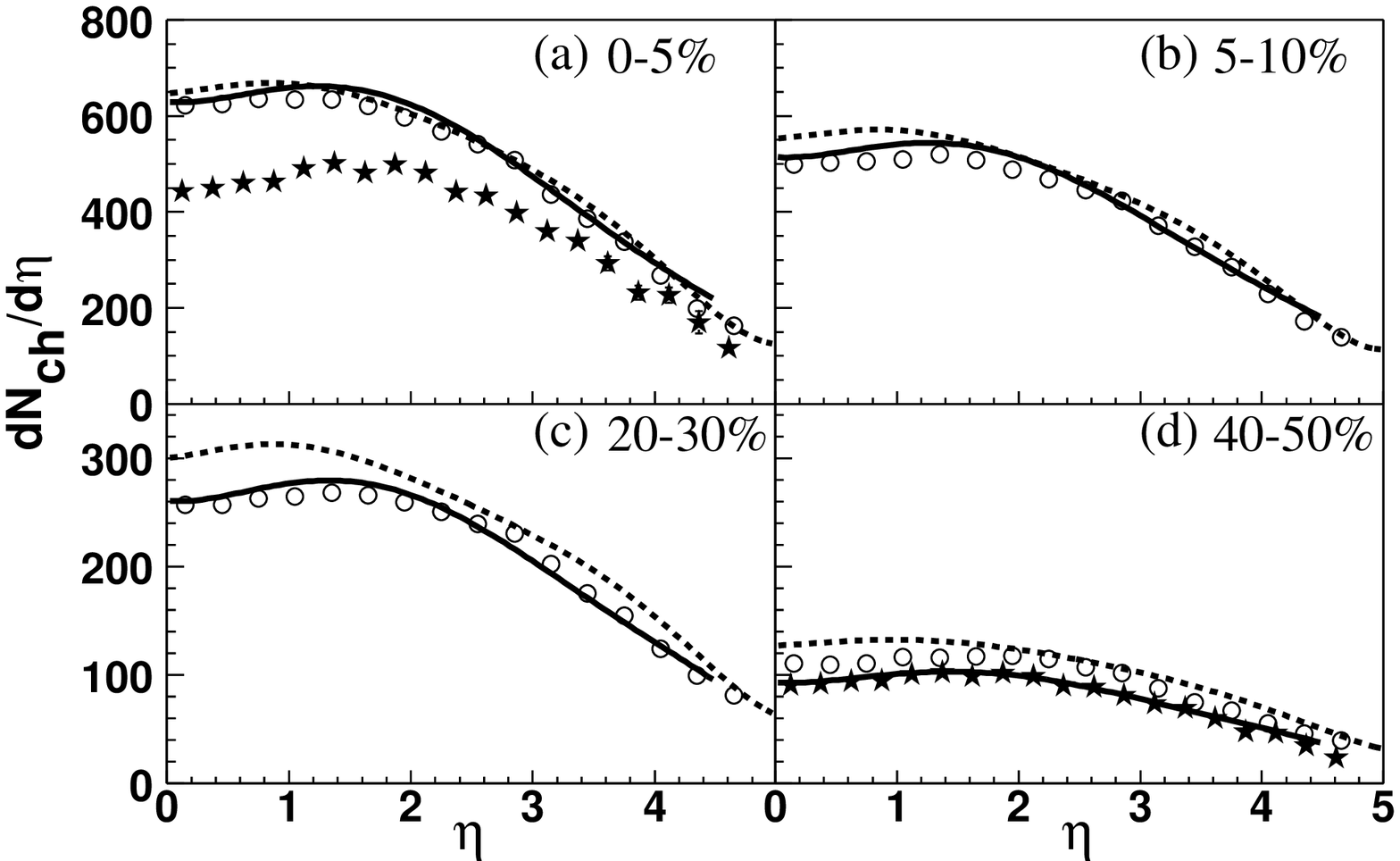}
}
\caption{The charged particle pseudorapidity distributions and 
Kharzeev-Levin saturation model fits. Left panel: PHOBOS 
data at 130~GeV~\protect\cite{KL130}. Right panel: BRAHMS data
at 200~GeV~\protect\cite{brahms200}.}
\label{fig:KLmultfits}
\end{figure}

It should also be noted that the saturation model was one of the
few models to correctly predict all of the following: the 130 and 200
GeV midrapidity multiplicity~\cite{PHO:mult,happylarry} and the
centrality dependence at all three energies~\cite{MDBQM2002}.
Figure~\ref{fig:KLmultfits} shows the fits to 130 and 200 GeV data
from PHOBOS and BRAHMS respectively.  The pseudorapidity and energy
dependence are primarily controlled by the $\lambda$ parameter, which
is extracted from deep inelastic scattering data.

So the initial state saturation model describes well the bulk of
soft particle production and, if pushed, may
also describe the moderately high $\pT$ particle production
behavior. More importantly, the hadron suppression or ``jet
quenching'' effect which we want to use as a probe of the density of
the strongly interacting bulk medium may not be a final state effect
at all, but may be actually be present in the gold wavefunction.

\subsection{Initial or Final State Effect?}

At the time of these lectures, RHIC was running deuteron-gold
collisions in order to resolve this issue. Initial state effects, such
as parton saturation, should still occur in \dAu\ since they are
associated with the gold nucleus itself and not the collision. Final
state effects, such as jet quenching, should go away in \dAu\ since we
do not expect a large bulk medium to form.  Some preliminary hints
already indicated that the suppression was probably a final state
effect rather than an initial state effect.

Since charm quarks are primarily formed by gluon-gluon fusion and are
not expected to be quenched in the final state~\cite{Dimacone}, charm
serves as a measure of the number of gluons available for hard
scattering from the initial state.  Open charm production, which was
found to scale with the number of collisions~\cite{Phxopenc}, implies
that parton saturation does not affect hard scattering.

\begin{figure}[htbp]
\centerline{
\epsfxsize=8cm\epsfbox{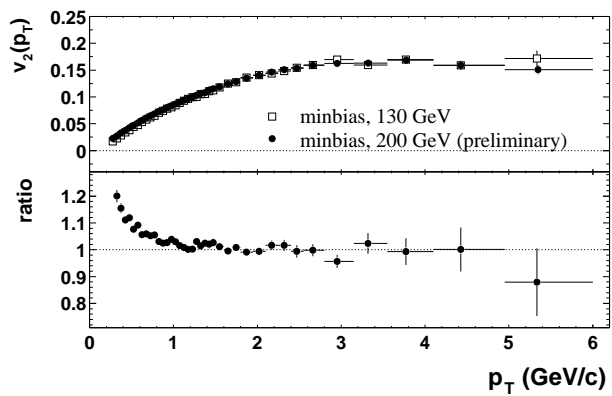}}
\caption{Elliptic flow of charged particles in 130 and 200 GeV \AuAu\
collisions (and their ratio) from STAR~\protect\cite{star:v2high}.}
\label{fig:v2high}
\end{figure}

The behavior of elliptic flow at high $\pT$ also suggests that high
$\pT$ particles are strongly absorbed in the final state.
Figure~\ref{fig:v2high} shows that elliptic flow reaches a constant
value at high $\pT$, independent of $\sqrtsnn$. Furthermore, the value
is so large that it is essentially the maximum allowable asymmetry
from a geometric point of view~\cite{VoloshinQM02}. This implies that
only jets emitted close to the surface make it out as was also
indicated by the approximate $N_{part}$ scaling of high $\pT$
particles.  Since the transverse geometry of the collision is a final
state effect and not present in a single initial gold wavefunction,
high $\pT$ particles must be strongly absorbed or rescattered in the
final state, such that the collision geometry leaves its imprint on
the final state momentum distribution. It should be noted that a $v_2$
value of 0.17 implies that twice as many particles are emitted
in-plane as out-of-plane, a huge effect.  This effect has been shown
by STAR~\cite{star:v2high} to persist to $\pT>8$~GeV/c.

Taken together, the results on overall particle production and high $\pT$
particle production are very suggestive.  The system appears to be
made up of hydrodynamic bulk matter.  The system is opaque and
expanding explosively, probably in all three dimensions.  The
estimated energy density is much higher than that of the theoretical
transition.  There is a freezeout along a universal curve near the
theoretical transition.  There is a strong suppression of inclusive
high $\pT$ yields and back-to-back pairs and an azimuthal anisotropy
at high $\pT$.  The natural implication is that there is a large
parton energy loss and surface emission.

These results are tantalizing, but there are some caveats. First of
all, we not yet have a complete 3D hydrodynamic description of the
collision which is consistent with all of the data. Additionally,
there are some outstanding puzzles from PHOBOS and PHENIX. Finally,
data from \dAu\ collisions are needed to really disentangle initial
state effects.  We will turn to the puzzles and \dAu\ data next.

\section{Some Puzzles at RHIC}

In addition to the surprising features mentioned above (blackness and
3D explosiveness of the source), there are two deep puzzles in the
data: the behavior of protons at moderately high $\pT$ and the 
apparent universality of particle production at high $\sqrtsnn$.

\subsection{Scaling Puzzle I: Baryon/Meson Differences}

\begin{figure}[htbp]
\centerline{
\epsfxsize=11cm\epsfbox{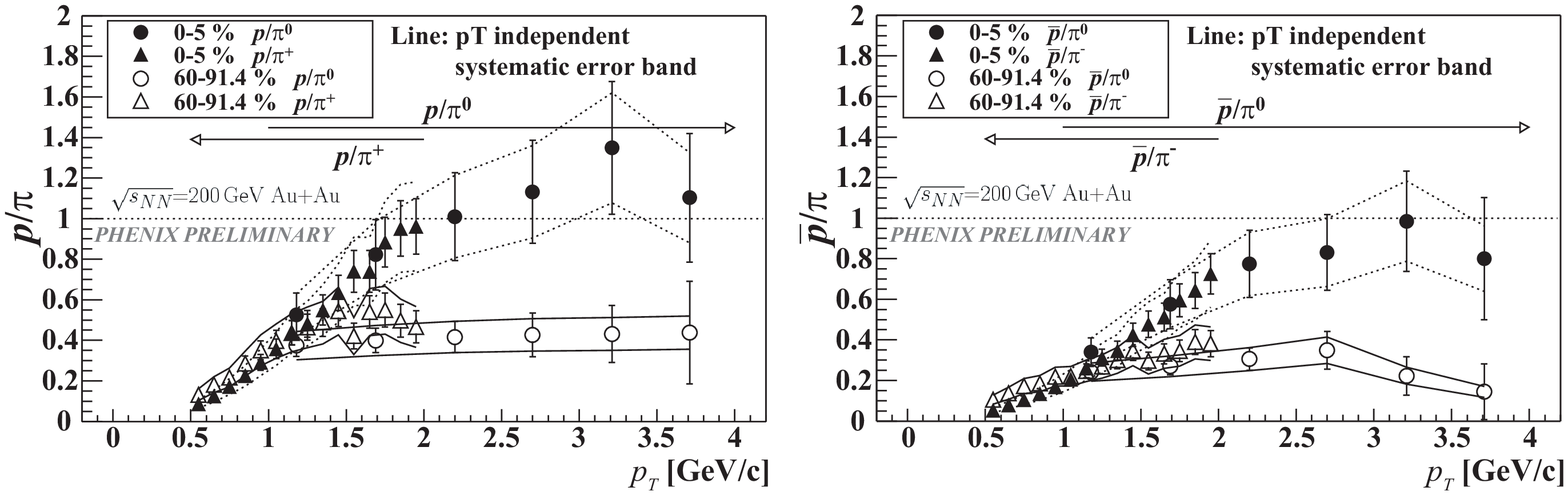}}
\caption{Proton to pion ratio for 200~GeV \AuAu\
collisions from PHENIX~\protect\cite{phx:ppi}.}
\label{fig:phxppi}
\end{figure}

The first puzzle, emphasized initially by PHENIX, concerns the
remarkable number of protons (compared to pions) at large transverse
momentum, as shown in Fig.~\ref{fig:phxppi}.  PHENIX has also
shown that pions are more suppressed than protons in the intermediate
$\pT$ region from 2--5 GeV/c~\cite{phxppiquench}. This effect is also
seen in neutral mesons and baryons by STAR~\cite{star:coal}.

Why are pions more suppressed than protons? The current ideas include
a modification to the fragmentation function in the hot medium or a
difference in gluon jet and quark jet quenching in the hot
medium. Perhaps the most intriguing explanation is that, in the
presence of jet quenching, a different production mechanism --- quark
coalescence --- starts to dominate hadron production.  Instead of the
usual jet fragmentation, this is a multiparton mechanism: three
independent quarks coalesce into a baryon or an independent quark and
antiquark coalesce into a meson~\cite{bm:coal}.

\begin{figure}[htbp]
\centerline{
\epsfxsize=8cm\epsfbox{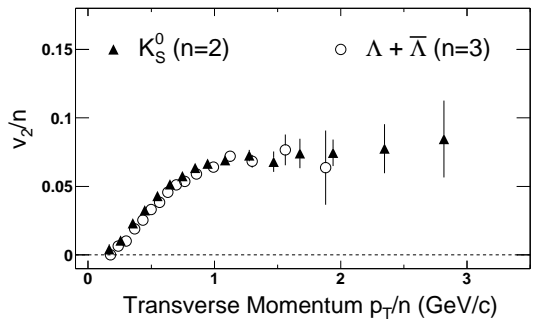}}
\caption{Elliptic flow per constituent quark as a function of
transverse momentum per constituent quark for lambdas and neutral
kaons for 200~GeV \AuAu~\protect\cite{star:coal}.}
\label{fig:starcoalv2}
\end{figure}

Since
$
(1 + 2 v_2 \cos 2\phi)^N \approx (1 + 2 N v_2 \cos 2\phi),
$
the coalescence model~\cite{bm:coal} predicts a scaling in elliptic
flow per constituent quark versus $\pT$ per constituent quark.
Figure~\ref{fig:starcoalv2}, from STAR, shows this scaling effect for
elliptic flow.  This model also explains the fact that, in
\AuAu\ collisions at high $\pT$, baryons and mesons behave similarly
while mesons are suppressed (and reach maximum $v_2$) at lower
momentum.

While this explanation is intriguing, this result remains a puzzle because
it is unclear that this model should apply to \dAu\ data (see 
Section~\ref{sec:dAu}).

\subsection{Scaling Puzzle II: Similarity of AA and $e^+e^-$ at High Energy}

\begin{figure}[htbp]
\centerline{
\epsfxsize=9cm\epsfbox{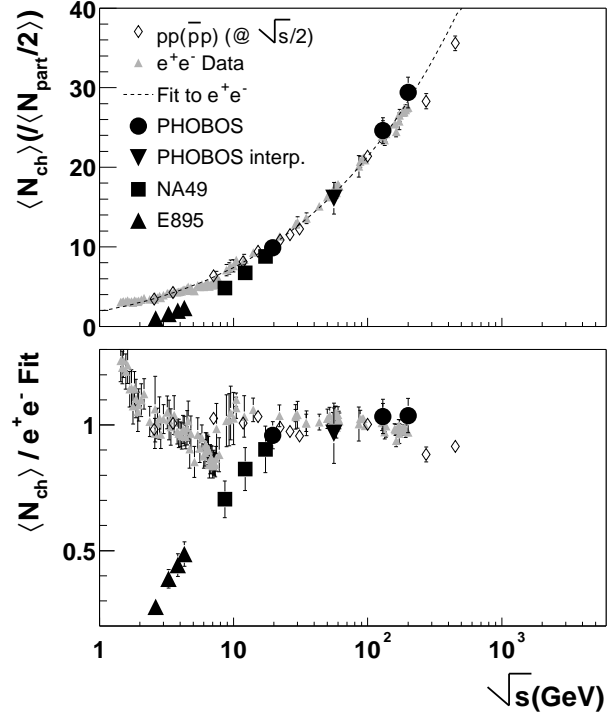}}
\caption{Comparison of the total charged multiplicity versus collision
energies for AA, \ee, \pp, and \ppbar\ data, as described in the text,
from PHOBOS~\protect\cite{phouniversal}. In the upper panel, the curve is a
perturbative QCD expression fit to the \ee\ data. In the lower panel,
the data have all been divided by the \ee\ fit.}
\label{fig:AAee}
\end{figure}

Figure~\ref{fig:AAee} shows the total charged multiplicity for AA
collisions (scaled by $\avgNp/2$) compared to \pp, \ppbar, and \ee, as
a function of the appropriate $\sqrt{s}$ for each
system~\cite{phouniversal}.  The \ee\ data serve as a reference,
describing the behavior of a simple color dipole system with a large
$\sqrt{s}$. The curve is a description of the \ee\ data, given by the
functional form: $C \alpha_{s}(s)^A e^{\sqrt{B/\alpha_{s}(s)}}$ with
the parameters A and B calculable in perturbative QCD and the constant
parameter C determined by a fit to the \ee\ data~\cite{Mueller}. In
order to compare them with \ee, the \pp\ and \ppbar\ data were plotted
at an effective energy $\sqrt{s_{\rm eff}}=\sqrt{s}/2$, which accounts
for the leading particle effect~\cite{Basile}. Finally, central AA
collisions, \AuAu\ from the AGS and RHIC, and \PbPb\ from CERN are
shown. Over the available range of RHIC energies from 19.6 to 200~GeV,
the \AuAu\ results are consistent with the \ee\ results, suggesting a
universality of particle production at high energy. In addition, the
\AuAu\ data approximately agrees with the scaled \pp\ and \ppbar\ data
suggesting that the effective energy of a high energy AA collision is
just $\sim \sqrtsnn$. This result is not understood theoretically and
remains a puzzle.

\section{The Latest Results from RHIC}\label{sec:dAu}

At the time of the lectures, the critical \dAu\ ``control'' run at
RHIC was not complete. Since then, results from this run have been
published by all four
collaborations~\cite{Bra:Raa200,Phx:dAu,Pho:dAu,Star:dAu}. These
results show no hadron or jet suppression in \dAu\, implying that the
suppression is NOT present in the nuclear wavefunction. This strongly
favors the jet quenching interpretation for hadron suppression in
\AuAu\ and has led to a lot of theoretical activity.

\begin{figure}[htbp]
\centerline{ 
\epsfxsize=49mm\epsfbox{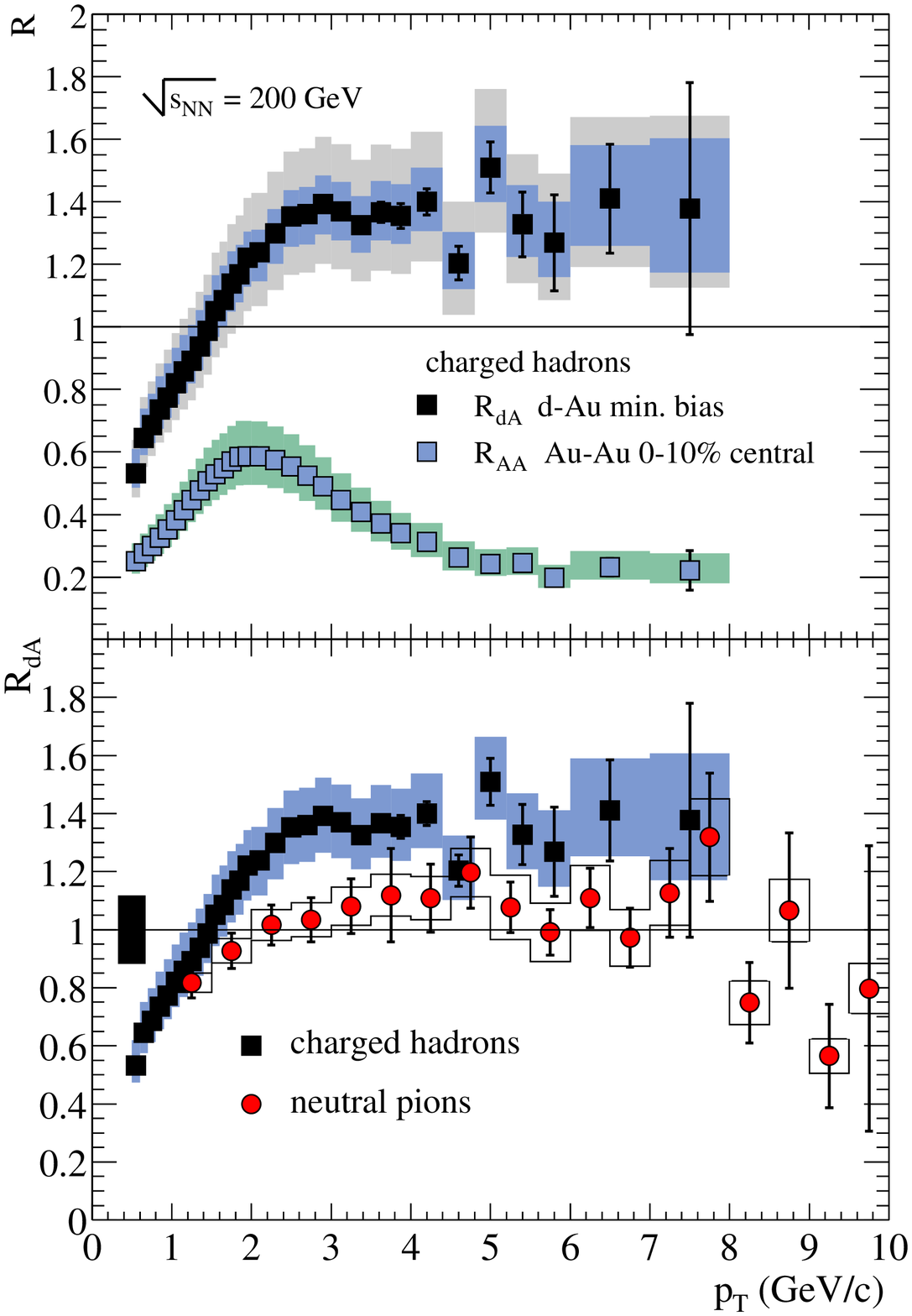}
\epsfxsize=65mm\epsfbox{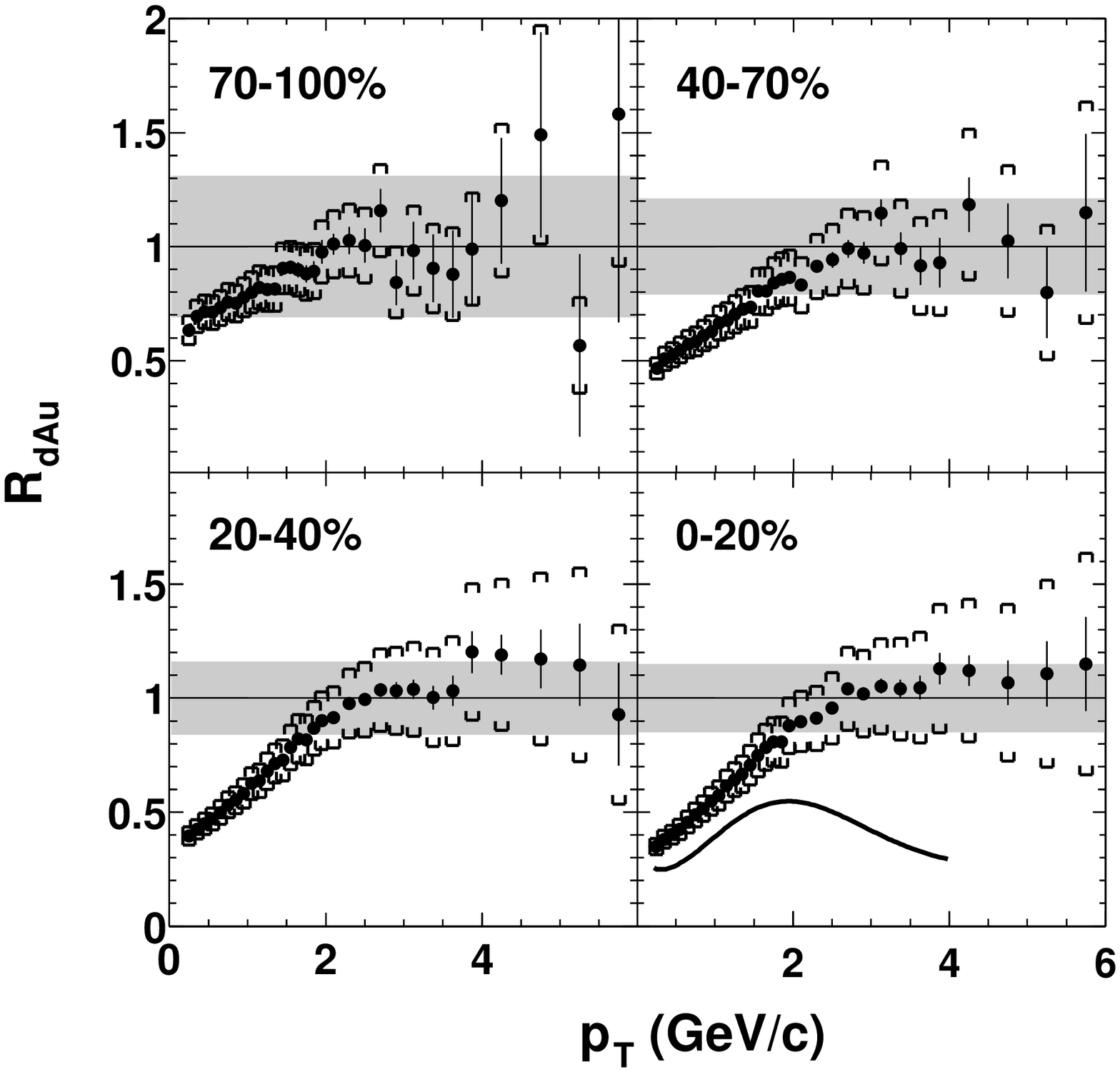} 
}
\caption{The nuclear modification factor. Left
panel: Midrapidity result from PHENIX. Minimum bias \dAu\ charged hadron 
result compared to central \AuAu\ charged hadrons and minimum bias \dAu\
neutral pions.~\protect\cite{Phx:dAu}. Right panel: PHOBOS results slightly
forward of mid-rapidity ($0.2<\eta<1.4$). The centrality
dependence of the nuclear modification factor for charged hadrons in
\dAu\ compared to central \AuAu~\protect\cite{Pho:dAu}.}
\label{fig:PhxPhodAu}
\end{figure}

Figure~\ref{fig:PhxPhodAu} (left panel) shows the minimum bias \dAu\
results from PHENIX at midrapidity. The charged hadrons are enhanced
rather than suppressed, in sharp contrast to \AuAu, a result confirmed
by BRAHMS~\cite{Bra:Raa200}.  Furthermore, the pions show collision
scaling, again in sharp contrast to the strong suppression seen in
\AuAu.

This contrast is striking, but the comparison of minimum bias \dAu\ to
central \AuAu\ is not fully decisive. Any nuclear effects in \dAu\ are
expected to manifest themselves primarily in central collisions and
can be washed out in minimum bias
collisions. Fig.~\ref{fig:PhxPhodAu} (right panel) shows the
centrality dependence of $R_{dAu}$ from PHOBOS, slightly forward of
midrapidity (from the deuteron's point of view). Even the most central
\dAu\ collisions show no suppression.

\begin{figure}[htbp]
\centerline{
\epsfxsize=8cm\epsfbox{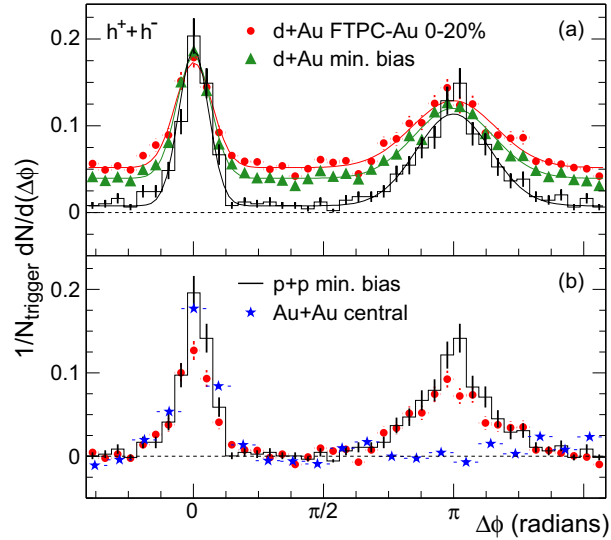}}
\caption{Azimuthal correlations for minimum bias \dAu, central \dAu,
and central \AuAu\ collisions compared to pp from
STAR~\protect\cite{Star:dAu}.}
\label{fig:StardAu}
\end{figure}

Finally, Fig.~\ref{fig:StardAu}, from STAR, shows that the jet
structure in central \dAu\ collisions can be understood based on a
\pp\ reference sample. There is no significant reduction
of back-to-back jets in head-on \dAu\ collisions.  The complete
suppression of the away-side jet in central \AuAu\ collisions is also
repeated in this plot for comparison.

Taken together, these results indicate that jets are quenched in
\AuAu\ collisions at RHIC energies ($\sqrtsnn$ of 130 and 200~GeV),
while there is little or no evidence of such quenching in \dAu\ or in
lower energy AA.

The production and behavior of protons in \dAu\ collisions is again
surprising, however.  Figure~\ref{fig:PhxPhodAu} shows that
midrapidity pions scale like $N_{coll}$ at high momentum while total
charged particles (including protons) are enhanced. This may explain
why PHOBOS (Fig.~\ref{fig:PhxPhodAu}) sees little enhancement of
charged particles (fewer protons for $\eta \sim 0.8$). The mystery
comes from the fact that the explanations put forward for the relative
behavior of protons and pions in \AuAu\ do not explain their behavior
in \dAu.

\section{Summary}

The field of heavy ion physics has indeed reached an exciting new era.
We have created a high temperature, high density,
strongly interacting bulk matter state in the laboratory, and we have
achieved temperatures higher than needed to theoretically create a
quark-gluon plasma. This bulk matter exhibits interesting properties.
It appears to be very dense and opaque even at high $\pT$, generating
the maximum possible elliptic flow and strongly quenching any jets
which are not formed on the surface of the material. Furthermore, the
system appears to be exploding in all three dimensions.

Some puzzles remain. Why are there so many protons at high $\pT$, and
why do protons and pions behave differently even in \dAu\ collisions?
Is the particle production universal between AA, \pp, \ppbar, and \ee\
at high energy, and if so, why?

Much work remains to be done to study this strongly interacting matter
more quantitatively and to resolve the puzzles. Fortunately, the
detectors and accelerator are undergoing continuous upgrades and the
prospects for a continued rich harvest of physics from RHIC look
excellent.

\section*{Acknowledgments}
Essential help in the assembly of this proceedings was provided by
David Hofman (U. Illinois, Chicago). Some of the material in the
original lecture presentation was provided by Barbara Jacak (SUNY,
Stony Brook) and Thomas Ullrich (BNL). This work was partially
supported by U.S. DOE grant DE-AC02-98CH10886.

\end{document}